\documentclass[graybox]{svmult}

\usepackage{mathptmx}       
\usepackage{helvet}         
\usepackage{courier}        
\usepackage{type1cm}        
\usepackage{makeidx}         
\usepackage{graphicx}        
\usepackage{multicol}        
\usepackage[bottom]{footmisc}

\makeindex             

\begin{document}

\title*{Exciton-Polariton Quantum Simulators}
\author{Na Young Kim and Yoshihisa Yamamoto}
\institute{Na Young Kim \at Edward L. Ginzton Laboratory, Stanford University, 348 Via Pueblo Mall, Stanford California, 94305 USA \email{nayoungstanford@gmail.com}
\and Yoshihisa Yamamoto \at Edward L. Ginzton Laboratory, Stanford University, 348 Via Pueblo Mall, Stanford California,94305 USA; National Institute of Informatics, 2-1-2 Hitotsubashi, Chiyoda-ku, Tokyo 101-8430, Japan}
%
%
\maketitle

\abstract*{A quantum simulator is a purposeful quantum machine that can address complex quantum problems in a controllable setting and an efficient manner. This chapter introduces a solid-state quantum simulator platform based on exciton-polaritons, which are hybrid light-matter quantum quasi-particles. We describe the physical realization of an exciton-polariton quantum simulator in semiconductor materials (hardware) and discuss a class of problems, which the exciton-polariton quantum simulators can address well (software). A current status of the experimental progress in building the quantum machine is reviewed, and potential applications are considered.}

\abstract{
\newline\indent A quantum simulator is a purposeful quantum machine that can address complex quantum problems in a controllable setting and an efficient manner. This chapter introduces a solid-state quantum simulator platform based on exciton-polaritons, which are hybrid light-matter quantum quasi-particles. We describe the physical realization of an exciton-polariton quantum simulator in semiconductor materials (hardware) and discuss a class of problems, which the exciton-polariton quantum simulators can address well (software). A current status of the experimental progress in building the quantum machine is reviewed, and potential applications are considered.}

\section{Introduction} 
\label{sec:1}

\begin{figure}[t]
\begin{center}
\sidecaption[t]
\includegraphics[scale=.45]{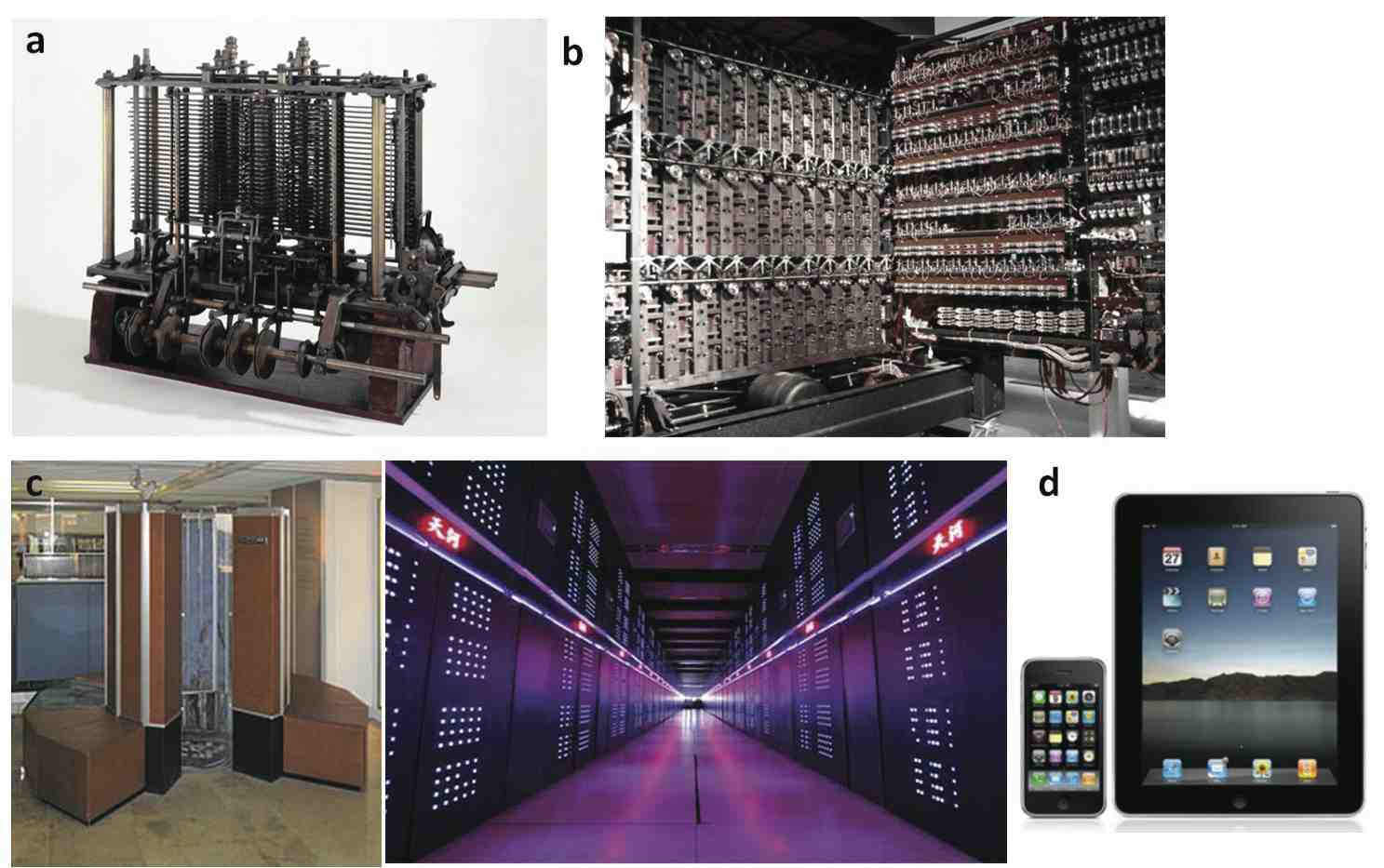}
\caption{Historical computing machines. \textbf{a}, The first fully automatic Babbage analytic calculator in the 19th century. \textbf{b}, An image of the Colossus, the first electronic digital Turing computer. \textbf{c}, Photos of the supercomputers, the first generation CRAY-1 built in 1970s (left), and Milkyway-2, the most powerful supercomputer in 2013. \textbf{d}, A multi-functional smart phone and a tablet computer.}
\label{fig:1}       
\end{center}
\end{figure}

We live in the twenty-first century -- the heart of the Information Age. A transition to the Information Age was greatly accelerated in the late 1990s by the rapid development of the technologies in digital computation, and short- and long-distance communications. Now computers are indispensable appliances in modern society, influencing all aspects of our lifestyle and activities. Classical computer hardware has been sophisticated as well as functionalized in terms of size, integrability and speed, pacing up with the advancement of electronics. Figure~\ref{fig:1} collects images of landmark machines throughout the computer history: a primitive Babbage automatic calculator, which could only compute simple arithmetics (Fig.~\ref{fig:1}a), a first digital Turing machine (Fig.~\ref{fig:1}b), which was programmable,  and cutting-edge supercomputers equipped with multiple processors (Fig.~\ref{fig:1}c) to palm-sized personal computers, which allow an individual to multitask actively and productively (Fig.~\ref{fig:1}d).  On top of the state-of-the-art computer platforms, myriads of software programs have also been designed and implemented successfully.

Notwithstanding that contemporary computers are impressively powerful and continuously improving, there still exist countless physical problems associated with many degrees of freedom, which demand unprecedented levels of performance in speed, flexibility and required resources beyond what the currently available computers can offer. Computer scientists theorized the computational complexity in regard to inherent difficulties of problem nature. Mechanically, this complexity may be quantified by computation time and resources for obtaining the right solution of a given problem. Quantum many-body problems belong to the category of the most complex class, that is often considered to be intractable by classical computers. This perplexing fact ignites the need  of novel computing machines for tackling such effortful challenges.

In 1982, Richard P. Feynman conceived a simple but brilliant idea, building a machine governed by quantum mechanical rules~\cite{Feynman82}. Hence, this quantum machine should be able to solve any quantum many-body problems~\cite{Mahan} by nature in an efficient and economical way. This approach is dubbed quantum simulation. This seed of idea is evolved to a universal quantum simulator by S. Llyod~\cite{Lloyd96}, which mimics a quantum system of interest through coherent operations on the simulating system.  Last decade, quantum simulation became a fast-growing theme in quantum science and technology both theoretically and experimentally. Several review articles have been published on the progress of quantum simulation research activities~\cite{Buluta09, Cirac12, Georgescu13}. 

\subsection{Digital and Analog Quantum Simulators}
\label{subsec:1}

The ultimate goal of the engineered quantum machine is to reach a universally acceptable explanation of unanswered phenomena~\cite{Buluta09, Cirac12, Georgescu13}. Often, physicists set up toy models, anticipating to explain the key features of the target phenomenon. Similarly, a quantum simulator aims a designated toy model, and it is prepared in two distinct ways: one is to build a bottom-up system with local addressing and individual manipulation, and the other is to directly map its time evolution of the system of interest onto the controlled time evolution of the simulating one~\cite{Buluta09, Cirac12, Georgescu13}. The former approach is named {\it{digital quantum simulator}} (DQS). In the DQS, quantum problems are tackled by arranging a large number of qubits, a unit of quantum information or a state vector,  followed by executing coherent unitary operations to manipulate these qubits. The universal DQS is an omnipotent quantum computer that can solve any quantum many-body problems efficiently~\cite{Ladd10}. The advantage of this scheme is the ability to supervise and identify errors during operations, which can be fixed by error-correction algorithms. A reliable and accurate solution can thusly be obtained. However, even simple prime number factorization problems require tremendous resources by the very nature, which is a major challenge to be overcome towards the implementation of a large-scale DQS~\cite{Georgescu13, Ladd10}.

On the other hand, the latter approach known as an {\it{analog quantum simulator}} (AQS) is free from such expensive restrictions since it only tries to resemble the system of interest as closely as possible. It does not need to define individual qubits and/or to perform error correction. In addition, the AQS is constructed for a specific problem not all quantum problems. Accordingly, the physical resources to build an AQS are much less required than those for a DQS. Therefore, scientists in many disciplines have started in establishing their own platforms of AQSs, combining currently available technology and benchmarking some famous toy models~\cite{Bloch12,Blatt12,Aspuru12,Houck12,Lu12}. Despite technical advantages of the AQS, there are fundamentally serious issues~\cite{Hauke12}: what specific problems are chosen to be simulated by the AQS?; and, what guarantees the reliability of the obtained solution to unknown problems with the AQS? In other words, it is meaningless to perfectly answer problems irrelevant to any real phenomena even if the AQS solves them accurately. Unfortunately, no methods to single out sources of errors in the AQS, not to mention no error correction protocols, are yet known. Hence, we all should remember  the underlying predicament of the AQS when we employ the AQS for significant contributions to our knowledge base. Keeping this in mind, in this section, we survey hardware and software of currently available AQSs.


\subsubsection{Hardware of the Analog Quantum Simulators}

\begin{figure}[t]
\sidecaption[t]
\includegraphics[scale=.5]{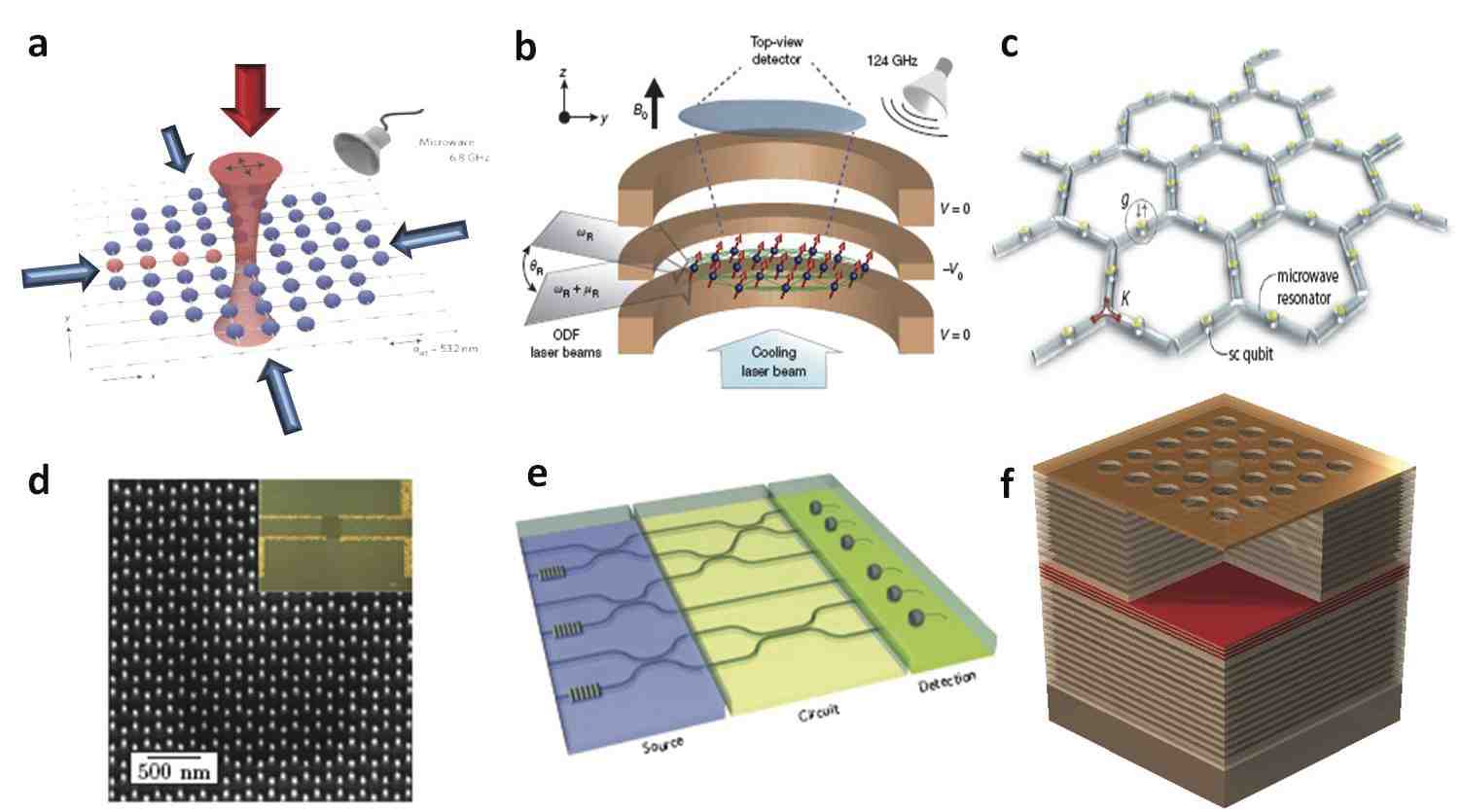}
\caption{Quantum Simulator Platforms. \textbf{a}, An illustration of ultracold atom gas in a two-dimensional optical lattice taken from Ref.~\cite{Bloch12}. \textbf{b}, A schematic of a two-dimensional triangular lattice formed by $^9$Be$^{+}$ ions in a Penning trap adapted from Ref.~\cite{Britton12}. \textbf{c}, A proposed microwave photon lattice using superconducting resonators and qubits in Ref.~\cite{Koch10}. Reprinted figure with permission from Koch et al., Phys. Rev. A 82, 043811 (2010). Copyright(2010) by the American Physical Society. \textbf{d}, Scanning electron microscopy image of a honeycomb lattice by nanofabrication in a two-dimensional electron gas system presented in Ref.~\cite{Simoni10}. \textbf{e}, A schematic picture of a planar photonic quantum simulator integrating essential components proposed in Ref.~\cite{Aspuru12}. \textbf{f}, An exciton-polariton quantum simulator in GaAs-based semiconductors. Permission of Figs (a, b, e) is acquired from Nature Publishing group.}
\label{fig:2}       
\end{figure}

The term `hardware'  here specifies physical elements of the AQSs akin to classical computer hardware. The AQS research activities were sparked by seminal work in 2002~\cite{Greiner02}, where I. Bloch and his colleagues observed a superfluid-insulator phase transition in ultracold Bose gases. This observation may be understood by the same physics of the metal-insulator transition in condensed-matter materials~\cite{Greiner02}. Assessing this successful demonstration, we generalize hardware of the AQS in three aspects: {\it{particles,  artificial lattices}} and {\it{detection schemes}}. For the last decade or so, the AQSs were built upon a variety of particles, diverse methods to engineer lattice geometries, and specialized probe techniques. Figure~\ref{fig:2} illustrates a few representative AQS platforms.

Current AQSs are based on ultracold bosonic~\cite{Bloch12} and fermionic~\cite{Esslinger10} atoms, trapped ions~\cite{Blatt12,Kimk11,Britton12}, electrons~\cite{Byrnes06, Byrnes07, Simoni10, Singha11}, superconducting qubits~\cite{Houck12, Koch10}, photons~\cite{Aspuru12,Angelakis07} and polaritons in a cavity~\cite{Hartmann06,Greentree06,Byrnes10,Na10}.  Many new particles or quasi-particles will likely join the list in the near future. Next, how are lattice potentials shaped for these particles? To trap cold quantum gases, pairs of oppositely propagating lasers are used to form spatial standing waves, periodic potentials in one-, two- and three-dimensions, also called optical lattices. This technique is superb in almost defect-free lattices and  the controllability of the potential amplitude by adjusting the intensity of the participating lasers. Responses of atoms to potential changes are recorded by time-of-flight absorption imaging, which maps the momentum distribution of particles. For example, the presence or absence of diffraction peaks as interferences distinguish  the coherent metallic phase from the incohrent insulating phase~\cite{Bloch12, Greiner02,Esslinger10}. Recently, a high-resolution real-space imaging technique was developed to perform a parity measurement of trapped atoms at a single site, which enabled quantification of correlation functions~\cite{Bloch12}. For charged ionic particles, a Paul trap was used to establish a string of ions~\cite{Kimk11}, and a Penning trap to arrange hundreds of ions in a triangular lattice~\cite{Britton12}. Typically the internal states of ions according to external manipulation are detected by fluorescence signals with photodetectors.  Among many techniques for confining electrons in semiconductors, the direct application of DC or AC electric and magnetic fields, strain fields for piezoelectric materials, and the direct etching processes are well established and controlled~\cite{Singha11}. DC and AC resistance, magnetoresistance and thermal resistances are quantities to characterize electrical transport properties under specific trapping potentials. Other particles, superconducting qubits, photons and polaritons share a common feature in light, and they are classified as photonic quantum simulators, a central topic of subsection 1.2.


\subsubsection{Software of the Analog Quantum Simulators}

\begin{figure}[t]
\sidecaption[t]
\includegraphics[scale=.5]{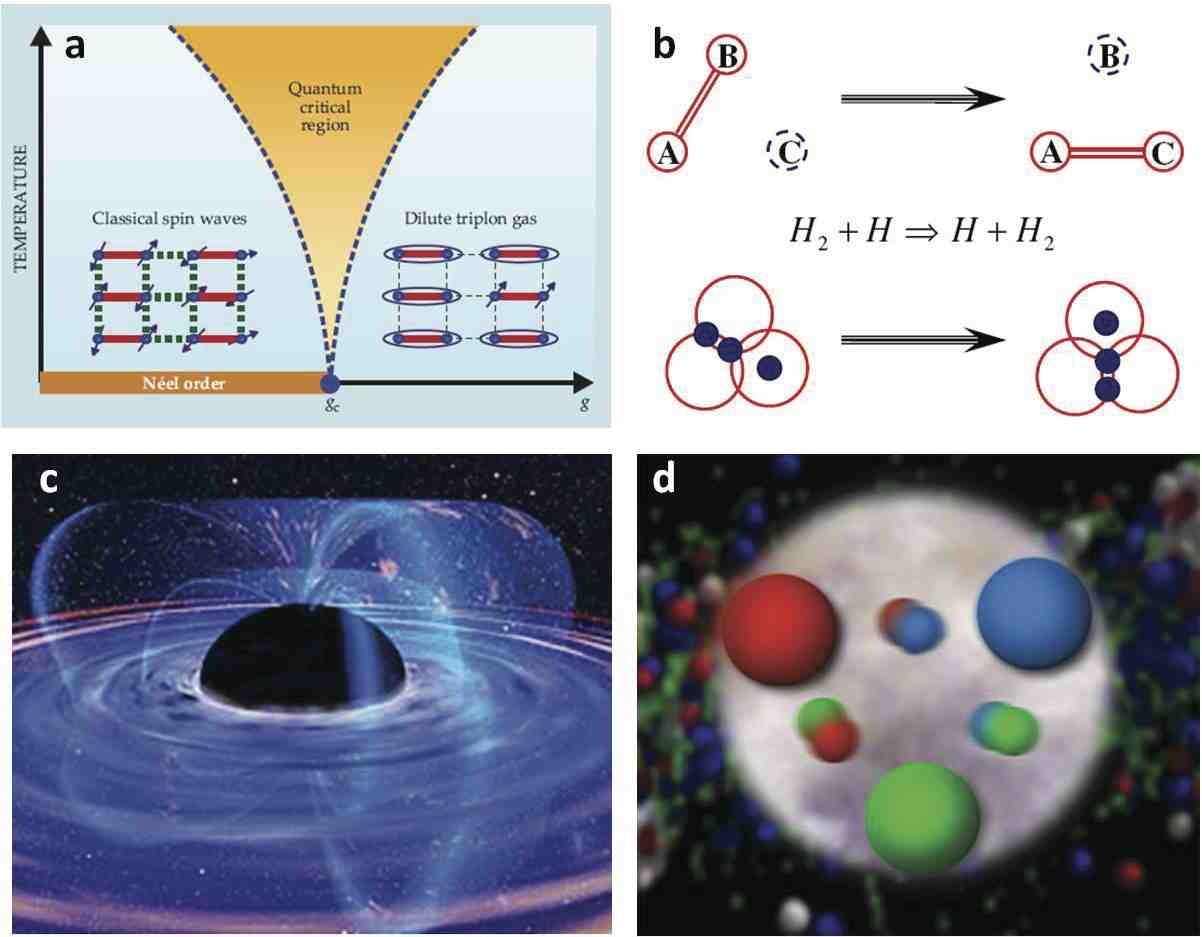}
\caption{Representative quantum many-body problems as possible applications of quantum simulators. \textbf{a}, A finite temperature phase diagram for the dimer antiferromagnet near the quantum critical point $g_c$ regimes adapted from Ref.~\cite{Sachdev11}. \textbf{b}, An exemplary hydrogen chemical reaction (top) to be simulated in coupled quantum dot systems (bottom) illustrated in Ref.~\cite{Smirnov07}. \textbf{c}, An artistic view of a Hawking radiation. \textbf{d}, A nucleon inner structure with quarks in quantum chromodynamics. }
\label{fig:3}       
\end{figure}

Let us turn to what software AQSs can run as applications. Early on, atomic AQSs studied a Hubbard Hamiltonian~\cite{Hubbard63}, one of the simplest but important toy models in condensed matter physics, for example, to exploit the electronic metal-insulator phase transition such as in transition-metal oxides~\cite{Tokura00}. The Hamiltonian consists of two energy terms to describe interacting particles in a lattice: a kinetic term $t$ for a nearest-neighbor hopping energy and an on-site Coulomb interaction energy term $U$. The Hamiltonian operator $\hat{H}$ is expressed in terms of particle operators $\hat{c}^\dagger, \hat{c}$ with a site index $i, j$, a spin index $\sigma$ and a particle density $\hat{n}=\hat{c}^\dagger \hat{c}$,

\begin{eqnarray*}
\hat{H} = -t \sum_{<i,j>,\sigma} (\hat{c}^\dagger_{i,\sigma} \hat{c}^{}_{j,\sigma}+h.c.) + U \sum^{N}_{i=1}\hat{n}_{i\uparrow}\hat{n}_{i \downarrow}, \nonumber
\end{eqnarray*}

where $<i,j>$ denotes the nearest-neighbor interaction on the lattice and $h.c.$ is the hermitian conjugate. Even though it is simple, the analytical solutions in higher dimensions than one dimension are not available, and obtaining solutions for even a small number of lattice sites in two dimension is numerically expensive.  Both Bose and Fermi-Hubbard Hamiltonians have been studied using bosonic~\cite{Greiner02, Bloch12} and fermionic~\cite{Esslinger10} gases and electrons~\cite{Singha11}.

The next one is an Ising Hamiltonian to study quantum magnetism dominantly favored in trapped ions~\cite{Kimk11, Britton12}. It is related to some of the challenging problems in condensed matter physics, such as the construction of a phase diagram of quantum matter and in particular quantum criticality as sketched in Fig.~\ref{fig:3}a~\cite{Sachdev08, Sachdev11}. Recently, stimulated by the discovery of new quantum matter arising from spin-orbital physics, tremendous efforts have been put into its simulation. Synthesizing the effective magnetic fields~\cite{Jaksch03, Lin09, Galitski13} to selectively manipulate the internal states of atoms in lattices provides a route to investigate the spin-orbital couplings in atom AQSs~\cite{Galitski13}. 

In quantum chemistry, AQSs will be useful to understand the static~\cite{Aspuru05} and dynamical~\cite{Kassal08} properties of molecules. Figure~\ref{fig:3}b depicts a proposal on how to implement the scattering process between a hydrogen atom and a hydrogen molecule using a coupled quantum dot system. The electron redistribution in three quantum dots under electrical gates would be anticipated to provide insights of the hydrogen dynamics~\cite{Smirnov07}.

Phenomena in cosmology and high-energy physics are often explored using gigantic apparatus due to the huge energy scales ($>$ keV) involved. However, the development of AQSs may grant an easy access to investigate such physics in tangible laboratory settings.  For instance, a Hawking radiation emitted from black holes (Fig.~\ref{fig:3}c) would be possibly explored with atomic superfluids~\cite{Giovanazzi05} or exciton-polaritons~\cite{Gerace12}. Relativistic Dirac physics has been addressed in a trapped ion~\cite{Gerritsma09}, and artificial Dirac points are created in fermionic atomic gases~\cite{Tarruell12} and copper-oxide molecules~\cite{Gomes11}.  In the future, quark physics inside the nucleus (Fig.~\ref{fig:3}d) would be an interesting quantum problem for AQSs. Since the aforementioned ones are a tiny subsection of quantum many-body problems in condensed matter physics, quantum chemistry, high-energy physics, cosmology, and nuclear physics, the phase space of AQS applications is almost boundless once reliable and functional AQSs are constructed.


\subsection{Photonic Analog Quantum Simulators}
\label{subsec:2}

Besides the many promising physical realizations of AQSs discussed earlier, this subsection specializes photonic AQSs, where light is a part of the particles. They include microwave photons coupled to superconducting qubits~\cite{Houck12,Koch10}, strongly interacting photons in assorted cavity shapes~\cite{Aspuru12,Angelakis07,Greentree06} and strongly coupled exciton-polaritons~\cite{Hartmann06,Byrnes10,Na10}. Most photonic systems would enjoy a planar structure, advantageous en route to large-scale architectures with the help of advanced nanofabrication processing techniques.  Photonic AQSs have also been proposed to reveal quantum phase transition of light~\cite{Angelakis07,Greentree06}. In addition, mapping the light polarization to a pseudo-spin `1/2' system allows to explore spin physics and magnetism. The unavoidable loss process through optical cavities makes photonic AQSs well suited to simulate open environment and non-equilibrium physics.

It is true that the exciton-polariton AQS also possesses an inherent photonic nature. However, the non-negligible matter nature from the excitons distinguishes the exciton-polariton AQS from other photonic AQSs. This particular chapter is devoted to this hybrid light-matter exciton-polariton AQS and begins by outlining both hardware and  software aspects in section 2 and section 3, respectively. Section 2 explains the fundamentals of exciton-polaritons and their condensation properties followed by technical details of lattice formation and experimental probes.  Potential physical problems the exciton-polariton AQS could target to address will be discussed in section 3. We give an overview of our experimental progress in constructing an exciton-polariton AQS, and perspective remarks in section 4.



\section{Hardware of Exciton-Polariton Quantum Simulators}
\label{sec:2}

We review the physical components of the exciton-polariton quantum simulator hardware: {\it{microcavity exciton-polaritons}} as our particle, methods to engineer {\it{periodic potential landscapes}} for exciton-polaritons  and {\it{microphotoluminescence}} imaging and spectroscopy as detection schemes.

\subsection{Particle:Exciton-Polariton Condensates}
\label{subsec:1}

\subsubsection{Microcavity Exciton-Polariton}

\begin{figure}[ht]
\sidecaption[t]
\includegraphics[scale=0.38]{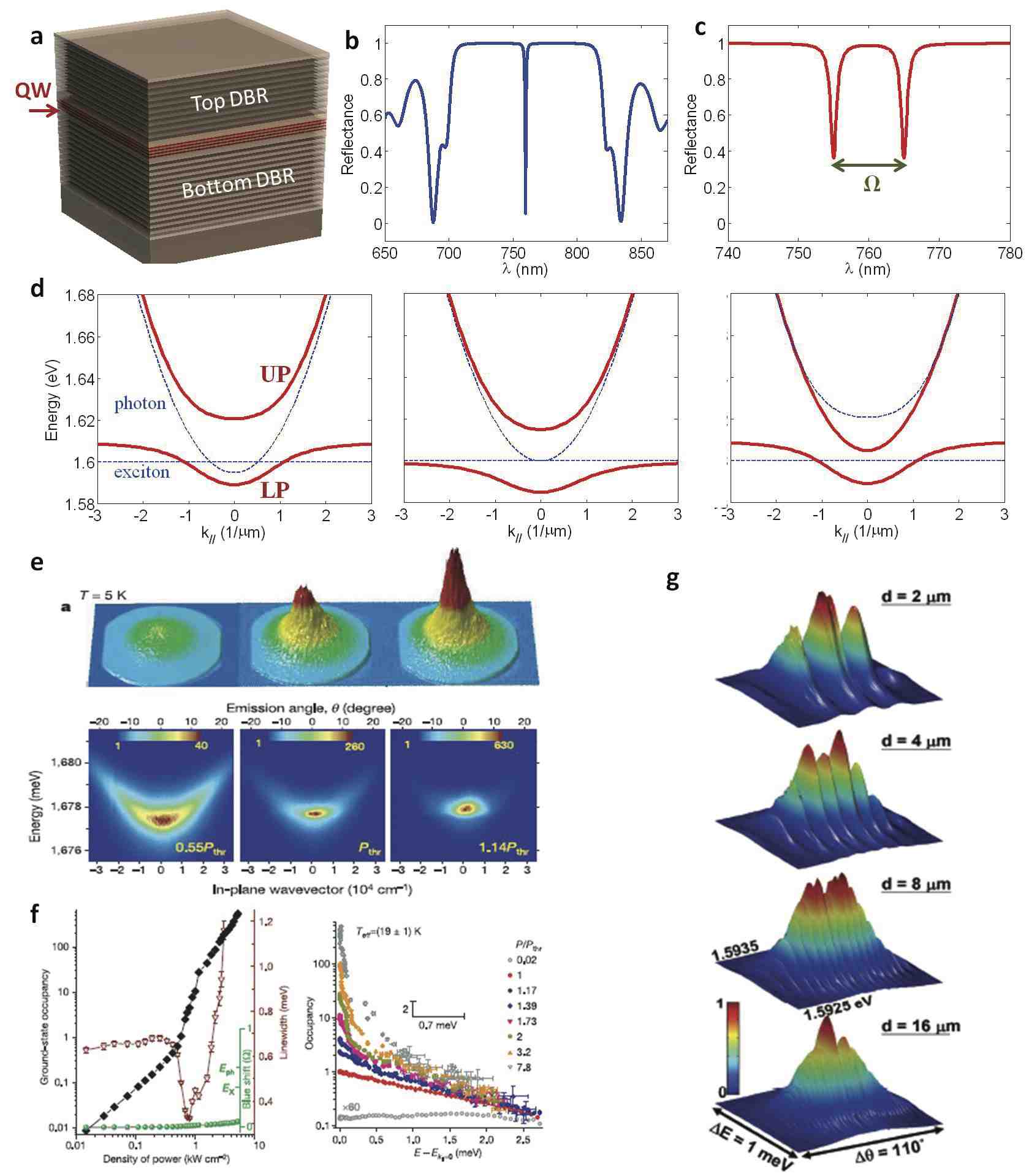}
\caption{Microcavity exciton-polaritons. \textbf{a}, A monolithically grown microcavity structure where multi-quantum wells (red layers) are embedded. textbf{b},  A calculated reflectance spectrum of a $\lambda$-sized cavity surrounded by pairs of a distributed Bragg mirror. textbf{c} A reflectance spectrum of two microcavity exciton-polaritons, upper polariton (UP) and lower polariton (LP), resulting from the strong coupling between a cavity photon and a quantum well(QW) exciton (\textbf{c}). The energy separation of the double dips  is the direct measure of the coupling strength between the photon and exciton denoted as $\Omega$. \textbf{d}, The energy dispersions of three configurations to mix the photon and the QW exciton modes, more photon-like LP mode (left, red-detuned), half-light and half-matter polaritons (middle, zero-detuned), and more exciton-like LP (right, blue-detuned). \textbf{e-g}, Experimental signatures of exciton-polariton condensations. \textbf{e}, Pump-power dependent LP population images (top) and spectra (bottom) in the momentum space and \textbf{f}, LP ground-state occupancy as a function of pump power (left) and energy (right) taken from Ref.~\cite{Kasprzak06}. \textbf{g}, Interference of LP condensates to exhibit spatial coherence from Young's double slit measurements reportd in Ref.~\cite{Lai07}. Nature granted permission of Figs. (e,f) from Ref.~\cite{Kasprzak06} and Fig.(g) from Ref.~\cite{Lai07}.}
\label{fig:4}       
\end{figure}

A `polariton' is a general term for a quasipatricle as an admixture of a photon and a certain excitation.  An exciton-polariton refers in particular to a resulting quasiparticle when a photon is mixed with an exciton in a semiconductor. Here, we consider exciton-polaritons in a two-dimensional (2D) microcavity structure with embedded quantum-wells (QWs) sketched in Fig.~\ref{fig:4}a. A photon is confined in a cavity and an exciton resides in the QWs. In 1992, C. Weisbuch {\it{et al.}} reported the first observation of photon-exciton coupled modes from a $\lambda$-cavity, where a single GaAs QW sits in the middle of the cavity~\cite{Weisbuch92}.  Since its discovery, the microcavity exciton-polariton system has been an attractive solid-state domain to investigate quantum boson statistical effects  such as condensation and superfluidity as well as to develop novel devices like polariton lasers or ultrafast spin switches both in theory and experiments~\cite{Kavokin,Snoke10,Deng10}. In this section, we summarize fundamentals of exciton-polaritons and their condensation properties. 

\paragraph{\textbf{Wannier Exciton}} %

Semiconductors are a class of materials with a gapped electronic excitation.  Numerous elemental and compound semiconductors have been the backbone of revolutionary electronic, optical and optoelectronic devices in the 20th century. Atoms are organized in a specific crystal structure, and the electrons of these atoms are brought together to form energy bands. In thermal equilibrium, all electrons are in the valence band, which is often called the system ground state or vacuum state. This stable low energy state is separated from the above excited states by an energy gap, and it is electrically insulating and optically dark. Suppose light shines onto a semiconductor. When the absorbed photon energy is bigger than the gap energy, it can promote electrons from the valence band to the conduction band, leaving positively-charged quasi-particles (holes) behind in the valence band. Similar to the hydrogen atom, where an electron is bound to a nucleus by Coulomb attraction, in perturbed semiconductors, an electron and a hole attract each other, creating an exciton, a Coulombically bound lower energy state~\cite{Yu}. 

This hydrogen-like entity is parameterized by a Bohr radius $a_B$, the extent of the electron-hole ($e-h$) pair wavefunction and a binding (or Rydberg) energy $E_B$, an energy cost to dissociate the pair. The complicated many-body system is simplified to a single particle picture by introducing an effective mass and a medium dielectric constant. Effective masses of electrons and holes in semiconductors are smaller than the bare electron mass, and the dielectric constant can be as large as 13 times of the vacuum case.  Therefore, compared to he hydrogen atom ($E_B = 13.6 $ eV, $a_B$ = 0.5 $\AA$), a  semiconductor exciton has a larger $a_B$ on the order of 0.1-10 nm and much weaker $E_B$ in the range of a few meV to 1 eV.  In large-dielectric-constant inorganic semiconductors like Si, GaAs, CdTe or ZnSe, the electric field is noticeably screened by the surrounding charges, yielding the weaker Coulomb interaction between electrons and holes. As a result, excitons are spread over many lattice sites, and are classified as `Wannier excitons'. Since the exciton is a primary excitation of semiconductors, it lives for a finite time and the system eventually returns to its equilibrium vacuum state. Typically, GaAs QW excitons have $a_B \sim$ 6-10 nm, $E_B \sim$10 meV, and lifetimes of 100 ps-1 ns~\cite{Tassone96}.

Composed of two fermions, electron and hole, an exciton behaves as a composite boson at low temperatures and in the low-density limit ($n_X$), where exciton wavefunctions do not easily overlap each other in a given volume. In 2D, the regime is marked by a density $n_X a_B^2  \ll 1$. Excitons in this dilute density regime obey Bose-Einstein statistics, and are expected to be condensed in the system ground state~\cite{Hanamura77,Griffin}.


\paragraph{\textbf{Cavity Photon}} %

In semiconductors, alternating $\lambda/4$-thick layers with different refractive indices makes a distributed Bragg reflector (DBR), a dielectric mirror.  Reflectance of the DBR increases with the number of the layer pairs and the refractive index contrast. At the interface of two layers, photons near wavelength $\lambda$ will reflect and constructively interfere from all interfaces. The simplest but useful resonator is a Fabry-Perot (FP) planar cavity, where photons are bounced back and forth between the DBR pairs. A $\lambda$-sized optical cavity would support only one longitudinal cavity mode for a sub-micron $\lambda$ value. Such a structure appears as a reflectance dip at the particular photon frequency of the designed cavity mode in broad-band reflectance spectroscopy (Fig.~\ref{fig:4}b). One figure of merit to describe a cavity is the quality-factor ($Q$), which relates to how long photons reside inside the cavity. The reported $Q$-factors of an empty FP DBR cavity ranges from a few thousands upto a million, corresponding to cavity photon lifetimes around  1-100 ps.

It is worth mentioning that a cavity photon would have an `effective mass' $m_c$ in the transverse plane. The growth direction confinement imposes that the longitudinal component $k_{\perp}$ of the light wave vector is given by $2\pi/\lambda$, which is much bigger than the transverse (or in-plane) wavenumbers $k_{\parallel}$. Therefore, in the regime of $k_{\parallel} \ll k_{\perp}$, the cavity photon energy $E_c$ can be approximated as follows:
\begin{eqnarray*}
E_c = \hbar v |k| & = \hbar v \sqrt{k_{\perp}^2+k_{\parallel}^2} = \hbar v k_{\perp} \sqrt{1+\left(\frac{k_{\parallel}}{k_{\perp}}\right)^2}, \nonumber\\
 &  \sim \hbar v k_{\perp} \left(1+ \frac{k_{\parallel}^2}{2k_{\perp}^2}\right)  \equiv E_{c0} + \frac{\hbar^2k_{\parallel}^2}{2m_c},\nonumber\\
\end{eqnarray*}
where $\hbar$ is the Planck's constant divided by 2$\pi$,  $v$ is the velocity in the medium, $E_{c0} = \hbar v k_{\perp}$, and $m_c$ satisfies the following relation $m_c = h/v\lambda$. In the visible to near infrared regions, $m_c$ values around $10^{-5} m_e$, and the energy dispersion ($E_c-k_{\parallel}$ relation) becomes parabolic with a stiff curvature due to extremely light effective mass.

\paragraph{\textbf{Exciton-Polaritons}} %

Consider a two-level single atom in a high-$Q$-cavity, which is one of the well-known problems in the cavity quantum electrodynamics. When the atomic transition energy is precisely on resonance with the cavity mode energy, the coherent energy transfer between the atom and the photon occurs reversibly at the  ``Rabi frequency" provided that the atom and cavity decay rates are much slower than the energy exchange rate. Spectroscopically, the strongly mixed modes appear as  an anti-crossing energy  gap known as ``vacuum Rabi splitting", a manifestation of $strong coupling$ between two modes.

Let us now turn into our case, a QW placed at the maximum of the photon field inside a high-$Q$ DBR cavity. A photon is absorbed to excite an exciton in the QW, and a photon is re-created inside the cavity resulting from the radiative recombination of the $e-h$ pair in the exciton, which can excite an exciton in turn. Such reversible and coherent energy transfer between photon and exciton modes continues within the cavity lifetimes. When this coupling occurs much faster than any other decay process in the system, the QW-cavity system also enters into a $strong coupling$ regime. And there emerge new photon-dressed excitons, which we name {\it{exciton-polaritons}} (or {\it{cavity polaritons}}). As a result, two anti-crossed energy branches, upper polariton (UP) and lower polariton (LP), appear and there exist two dips at UP and LP energy states separated by the interaction strength $\Omega$ in reflectance spectroscopy (Fig.~\ref{fig:4}c).
In a linear signal regime, the coupled-cavity photon-QW exciton Hamiltonian $\hat{H}$ is written by a cavity photon operator $\hat{a}_k$ with energy $\hbar \omega_{ph}$, a QW exciton operator $\hat{C}_k$ with $\hbar \omega_{exc}$ and their in-between interaction coupling constant $\Omega_k$,

\begin{eqnarray*}
\hat{H} =\hbar \sum_k [\omega_{ph}\hat{a}^\dagger_k\hat{a}_k + \omega_{exc}\hat{C}^\dagger_k\hat{C}_k - i\Omega_k(\hat{a_k}^\dagger\hat{C}_k-\hat{a}_k\hat{C}^\dagger_k)], \nonumber\\
\end{eqnarray*}
where $k$ is the simplified notation for $k_{\parallel}$. Introducing an exciton-polariton operator at a momentum $k$ as a linear superposition of the cavity photon and the QW exciton modes $\hat{P}_k = u_k\hat{C}_k + v_k\hat{a}_k$, the Hamiltonian is diagonalized, reaching a simplified Hamiltonian $\hat{H}_T$

\begin{eqnarray*}
\hat{H}_T = \sum_k \hbar \Omega_k \hat{P}_k^\dagger\hat{P_k}, \nonumber\\
\end{eqnarray*}
and the exciton-polariton energy dispersion is explicitly given by

\begin{eqnarray*}
\hbar\omega_k = \frac{1}{2}\left( \hbar(\omega_{exc}+\omega_{ph})+i\hbar(\gamma_{ph}+\gamma_{exc}) \right) \pm \frac{1}{2}\sqrt{(2\hbar\Omega_k)^2+(\hbar(\omega_{exc}-\omega_{ph})+i\hbar(\gamma_{exc}-\gamma_{ph}))^2}. \nonumber\\
\end{eqnarray*}
$\gamma_{ph}$ is the photon-mode decay rate through the cavity, and $\gamma_{exc}$ is the non-radiative decay rate of the exciton.

Figure~\ref{fig:4}d illustrates the energy ($E$) versus the transverse wavenumber ($k_{\parallel}$) of independent cavity photon and exciton modes in blue dashed lines and the UP and LP in red straight lines. Effective masses and lifetimes of UP and LP are determined as a weighted average of individual constituents, taking into account of photon and exciton fractions. Near $k_{\parallel}=0$, the photon-like LP branch has an extremely light effective mass, but it is exciton-like at large $k_{\parallel}$ values with a flatter dispersion, corresponding to a heavy mass. Depending on fractions of photons and excitons ($|u_k|^2, |v_k|^2$), three different detuning regimes (red, zero, blue) are possible as shown in Fig.~\ref{fig:4}d. A detuning parameter $\Delta$ is defined as $\Delta (k_{\parallel}=0) = E_c (k_{\parallel}=0)-E_{exc}(k_{\parallel}=0)$. For instance, the negative $\Delta$ is red detuned, where the photon fraction is bigger, whereas the positive $\Delta$ is blue-detuned with exciton dominance.

\subsubsection{Optical Processes: Radiative recombination vs Relaxation processes}

Before delving into the properties of exciton-polariton condensation in the following section, we give a simple picture of optical processes inside a QW-microcavity structure. As seen earlier, when the QW-microcavity is excited by an external pump, $e-h$ pairs are created. These high-energy particles are quickly cooled down to large-momentum excitons or exciton-like polaritons. The dynamics of exciton-polaritons can be primarily understood by the interplay between radiative recombination and energy relaxation processes. The former radiative recombination rate of exciton-polaritons at $k_{\parallel}$ is simply determined by the lifetime of photon and exciton modes with appropriate fractional values for each mode. Therefore, small $k_{\parallel}$ exciton-polaritons decay faster due to short-lived photons, whereas those of large $k_{\parallel}$, more exciton-like ones, have smaller radiative recombination rates, and hence a longer lifetime.

Two dominant exciton-polariton relaxation processes have been extensively studied: one is due to exciton-phonon interaction~\cite{Tassone96,Tassone97}, and the other due to exciton-exciton interaction~\cite{Tassone99,Porras02}. Tassone and his colleagues quantified polariton-acoustic phonon scattering rates in a microcavity with a single GaAs QW, which were calculated by the deformation potential interaction using a Fermi golden rule~\cite{Tassone96, Tassone97}. Setting up semiclassical Boltzmann rate equations, Tassone {\emph{et al.}} observed that large-momentum polaritons are efficiently relaxed to lower energy regions through the emission or absorption of  acoustic phonons, and that these scattering processes are  inelastic and incoherent. However, they also found that phonon scattering is significantly reduced towards low $k_{\parallel}$ values of the LP branch, where the LP effective mass becomes lighter and the LP density of states is smaller. Hence,  exciton-polaritons are rather parked at the deflection of the LP dispersion branch, the {\it{bottleneck}} effect.


To overcome the bottleneck effect, people have resorted to another important relaxation process induced by elastic exciton-exciton scattering in the 2D QW~\cite{Ciuti98}.  When two excitons encounter,  two nearby excitons interact via either direct  or exchange interactions among electrons and holes. The direct term comes from  dipole-dipole interactions of excitons, which are considered to be negligible in QW.  Up to first order, the dominant mechanism is short-ranged Coulomb exchange interactions between electron and electron or between hole and hole~\cite{Tassone99,Ciuti98}. An exchange interaction strength is roughly on the order of $6E_Ba_B^2/S$, where $S$ is the quantization area. Conserving energy and momentum, coherent scatterings are further amplified by the number of LP, which is the bosonic final-state stimulation effect. This process is a crucial mechanism to accumulate LPs near $k_{\parallel}=0$, reaching condensation.

\subsubsection{Exction-Polariton Condensation}

At low temperature and a dilute density, exciton-polaritons are also regarded as composite bosons, governed by Bose-Einstein statistics. Immediately after the discovery of the exciton-polaritons in the microcavity-QW structure~\cite{Weisbuch92}, people recognized the possibility of exciton-polariton condensation~\cite{Imamoglu96}. To validate exciton-polariton condensation at $k_{\parallel}=0$, we should clarify system conditions with care: first, the exciton-polariton density is low enough, not exceeding the Mott density, which describes the state where the concept of excitons breaks down due to particle screening. The Mott density is roughly set by 1/$a_B^2$ in 2D, and it is critical to maintain the GaAs exciton density per QW at less than 10$^{12}$ cm$^{-2}$. One way to achieve this is to increase the number of QWs in the cavity, decreasing the average exciton density per QW by the assumption that the created excitons are equally distributed over all QWs. We commonly place 4, 6, 8 or 12 GaAs QWs in a AlAs cavity for this reason.  Second, coherence among exciton-polaritons at $k_{\parallel}=0$ should occur spontaneously not from the coherence of the pump laser. A safe manner of the excitation scheme is to choose the excitation laser energy to be much higher than the polariton ground state ({\it{incoherent excitation}}), which injects either high energy $e-h$ pairs or excitons as we described in the previous subsection 2.1.2. Through incoherent phonon scattering processes, the initial energy and momentum of the injected particles are lost, and LPs are cooled down to the lowest energy and momentum state, establishing coherence through the polariton-polartion coherent scattering processes.  

A series of experimental evidence or signatures of exciton-polariton condensation near $k_{\parallel}=0$ was reported in GaAs, CdTe and GaN inorganic~\cite{Deng02,Kasprzak06,Deng06,Balili07,Deng07,Chrisopoulos07} and organic~\cite{Cohen10, Plumhof14} semiconductors. A first piece of evidence is the observation of macroscopic population at the system ground state. Figure~\ref{fig:4}e shows pump-power dependent LP population distribution in momentum space (top panel) and their spectroscopic images (bottom panel) in CdTe-based semiconductors reported by Kasprzak {\it{et al.}}~\cite{Kasprzak06}.  When the particle density reaches quantum degeneracy threshold, bosons attract to be in the same state,  thermodynamically favorable state enhanced by the boson stimulation effect~\cite{Deng02,Kasprzak06,Balili07}. Next, the essential physical concept of the condensation is off-diagonal long-range order related to spatial coherence in condensation~\cite{Kasprzak06,Deng07,Lai07}. As a system order parameter, the  complex-valued LP wavefunction $\psi(\vec{r},\phi)$ is  $\psi(\vec{r},\phi) = \sqrt{n(\vec{r})} \exp (-i\phi(\vec{r}))$  determined by two real values, the LP density $n(\vec{r})$ and  the real-space phase $\phi(\vec{r})$. By means of an interferometer, the phase information of the order parameter is obtained, and the spatial coherence has been measured ~\cite{Kasprzak06,Deng07}.  Section 2.1 gave an overview of basic mathematical description of exciton-polaritons, the primary optical processes to lead to condensation and its general characterization in a standard QW-microcavity structure. The next subsection will summarize methods of engineering lattices for exciton-polaritons.


\subsection{Lattice:Photonic and Excitonic Lattices}
\label{subsec:2}

The second aspect of the exciton-polariton quantum simulator hardware is how to impose artificial crystal lattices on exciton-polaritons. Exploiting the duality of exciton-polaritons emphasized in section 2.1, we are able to create the potential landscape for either photon or exciton modes. Within a 2D planar microcavity-QW structure, the engineered lattices lie in  zero-, one- or two-dimensions. In this section we describe various designs to create the lateral potential lattices attempted in the exciton-polariton systems.

\begin{figure}[t]
\sidecaption[t]
\includegraphics[scale=.45]{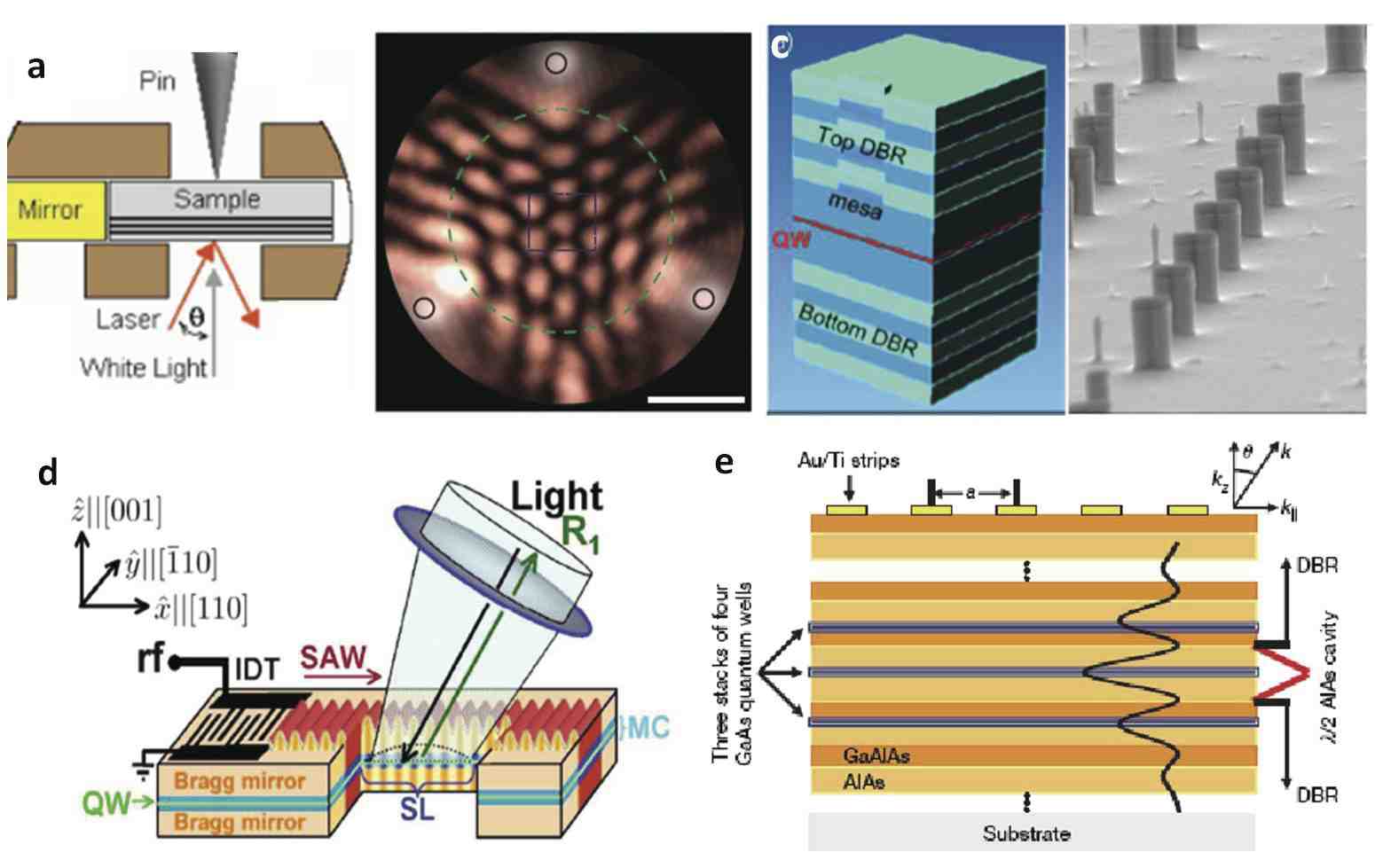}
\caption{Generations of trapping potentials in microcavity exciton-polaritons. \textbf{a}, Mechanical strain introduced by a pin near the sample changes the exciton energy (Ref.~\cite{Balili07}). \textbf{b}, A honeycomb lattice of exciton-polaritons induced by a three-spot laser pump profile Ref.~\cite{Tosi12}. \textbf{c}, A partial (left, Ref.~\cite{Nardin10}) and complete etching (right, Ref.~\cite{Galbiati12}) methods to introduce a photonic trap. Reprinted figures with permission from Nardin et al., Phys. Rev. B 82, 045304 (2010) (left), and Galbiati et al., Phys. Rev. Lett. 108, 126403 (2012) (right). Copyright(2010, 2012) by the American Physical Society. \textbf{d}, A dynamic phonon lattice in a piezoelectric GaAs microcavity-QW structure reported in Ref.~\cite{deLima06}. Reprinted figure with permission from de Lima et al., Phys. Rev. Lett. 97, 045501 (2006). Copyright(2006) by the American Physical Society. \textbf{e}, A weakly modulated in-plane one-dimensional photon lattice employing a thin-metal film technique on a grown wafer. A figure is adapted from Ref.~\cite{Lai07}.}
\label{fig:5}       
\end{figure}

\subsubsection{Methods of Creating Polation Lattices}

Figure~\ref{fig:5} collects a variety of methods to produce in-plane potentials for exciton-polaritons. We classify them into two groups, excitonic and photonic traps depending on which component is modified by the method. A mechanical stress induced by a sharp pin (Fig.~\ref{fig:5}a) shifts the QW exciton energy to a lower energy side (red-shifted) acting as an exciton trap~\cite{Balili06}. Its induced potential strength can be as large as hundreds of meV, and lateral sizes of the potential are determined primarily by tip sharpness. Because QWs are far below the top surface, QW excitons face a rather broadened potential, whose lateral size is on the order of  micrometer.  Applying electric fields is widely used to modify the QW exciton energy via the quantum-confined Stark effect~\cite{Miller84}. Magnetic fields shift the QW exciton energy as well.  Although AC electric and magnetic fields can be applied, to the best of our knowledge all of the above methods have so far been implemented using static sources.  Dynamical potentials were successfully launched by surface acoustic waves in piezoelectric GaAs semiconductors, which basically patterns propagating trap potentials~\cite{deLima06}. So far, 1D~\cite{Cerda10} and 2D~\cite{Cerda13} square lattices have been produced. This method still suffers from limited potential depth on the order of  hundreds of $\mu$eV and a small number of possible patterns according to the symmetry properties of the host material's piezoelectric tensor.

On the other hand, several clever techniques were devised for photon traps.  Chemical and dry etchings are common techniques for semiconductor processing.  Either partially etching the cavity layer~\cite{ElDaif06,Nardin10} or completely etching all layers, except for the designed area~\cite{Bloch98}(Fig.~\ref{fig:5}c), is done on microcavity-QW wafers. Partial cavity-layer etching method was developed in B. Deavaud's group in Switzerland, and resulted in tens of meV-strong potential, which quantizes cavity photon modes like zero-dimensional quantum-dot~\cite{ElDaif06,Nardin10}.  Recently, the complete etching method for making a single pillar was extended to make 1D wire ~\cite{Wertz10} and 2D pillar-array potentials~\cite{Jacqmin14}.  Despite recent progress, roughness of the sidewall surface has been the primary concern and caused degradation of the quality of QWs as well as spoiling the strong coupling regime especially for sub-micron sized pillars. Partial cavity-layer etching technique is in this sense less detrimental. Lately, several groups have noticed that the pump laser profile can be designed to trap exciton-polaritons through stimulated scattering gain under the pump profile~\cite{Roumpos10,Tosi12}. This technique can also be implemented to induce vortex lattices of exciton-polaritons due to the particle-particle repulsive interactions shown in Fig.~\ref{fig:5}b~\cite{Tosi12}.

Our group employs a thin metal-film technique to spatially manipulate cavity photon-mode energy illustrated in Fig.~\ref{fig:5}e~\cite{Lai07,Kim08}. Patterned by electron-beam lithography followed by a metal deposition on a grown wafer, a metal film modulates spatially the cavity lengths. The basic principle of such effect on the photon mode can be understood by looking at the boundary conditions of the electromagnetic (EM) fields at the interface between the top semiconductor layer and the metal film. Unlike bare surfaces of the semiconductor layer to vacuum or air where EM fields smoothly decay to  an open medium, EM fields are pinned to be zero at the metal-semiconductor interface. Consequently, EM fields are squeezed under the metal film, effectively shortening the cavity-layer thickness, which results in a higher photon energy. This method enjoys the simplicity and design flexibility, and it is completed as a post-processing step to the grown wafer, leaving the QWs intact. 1D~\cite{Lai07}  and 2D~\cite{Kim11,Masumoto12,Kim13,Kusudo13} geometries have been preprared. Despite the advantages, the generated potential strength is weak (100-400 $\mu$eV), and its actual lateral potential profile in the QW planes is broadened for the same reason as mentioned for the strain-induced potential described earlier~\cite{Kim08}. In all excitonic and photonic potentials, real potential strength for exciton-polaritons should be adjusted by the fraction of the modified component at different detuning values.

\begin{figure}[t]
\sidecaption[t]
\includegraphics[scale=.5]{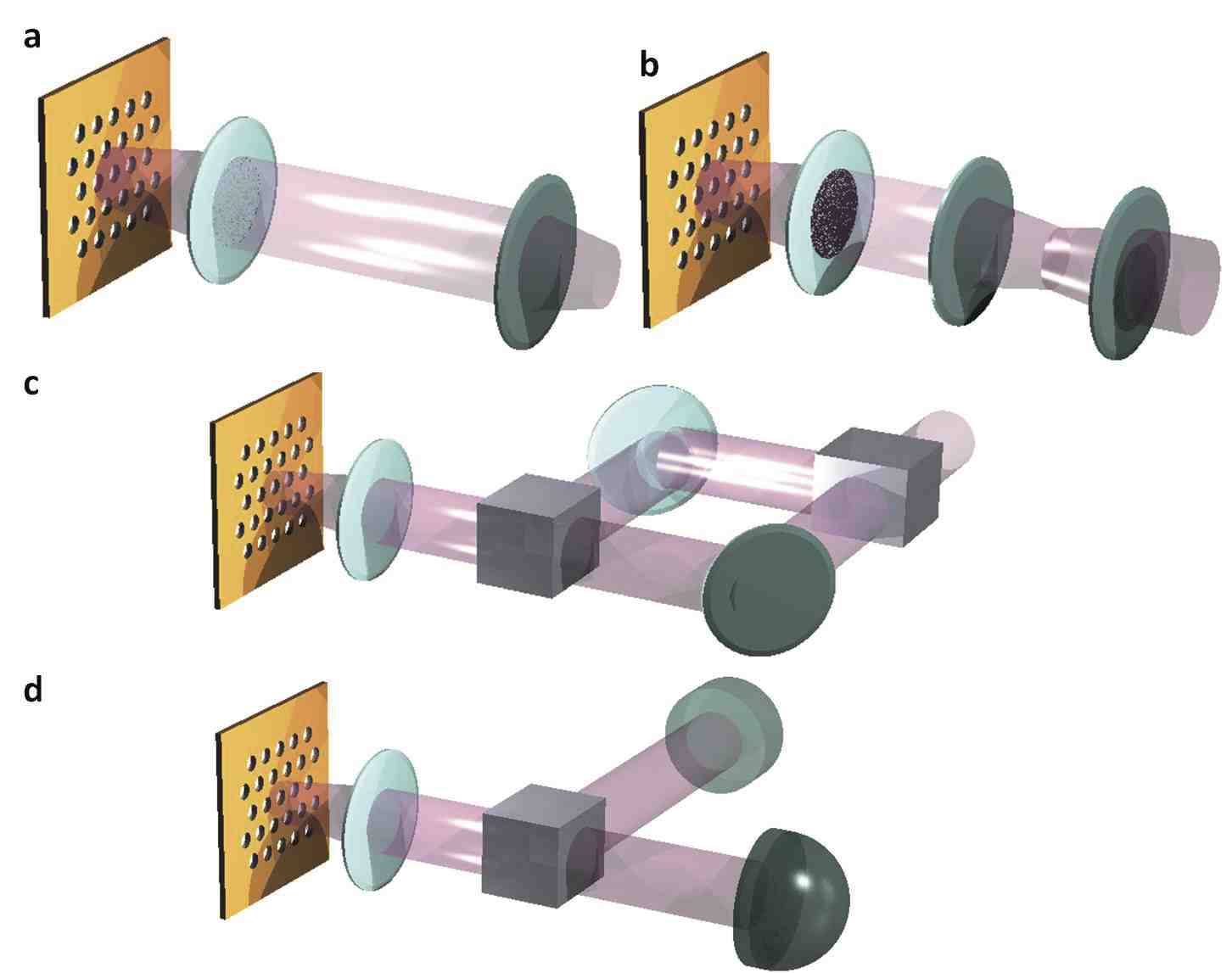}
\caption{Schematics of Fourier-optic measurement setups for exciton-polariton optical characterization. \textbf{a}, A real space configuration of the sample plane  mapped to an image plane of detectors. \textbf{b}, A Fourier transformed setup to access the momentum space information by inserting one more lens. \textbf{b}, A Mach-Zehnder interferometer to access phase in real-space. \textbf{d}, A Hanbury Brown-Twiss intensity correlator setup. }
\label{fig:6}       
\end{figure}

\subsection{Probes: Photoluminescence}
\label{subsec:3}

Now we turn to the third aspect of the hardware, the measurement schemes to identify optical properties of exciton-polaritons in the QW-microcavity structures.

\subsubsection{Photoluminscence Setup}

Optical properties of semiconductors are probed in various ways: absorption, reflection, transmission and luminescence through responses to optical excitations~\cite{Yu}.  Exciton-polariton characteristics are primarily examined via photoluminescence (PL). High energy light shines onto a QW-microcavity structure, creating exciton-polaritons which are eventually leaked out through the cavity in a form of photons. The energy, wavenumber (momentum), polarization of the exciton-polaritons are uniquely transferred to those of the leaked photons. Thus, by observing the leaked photons, we obtain energy, momentum and polarization information of the exciton-polaritons using a spectrometer and the temporal dynamics using a streak camera. Since the QW-exciton binding energy in GaAs and CdTe semiconductors is around 10 meV, experiments are performed in a cryostat held at low temperatures around 4-10 K, where thermal energy cannot dissociate the excitons. In GaN, ZnO or organic semiconductors whose exciton binding energy exceeds the room-temperature thermal energy $\sim$26 meV, their PL setup can be operated at room temperatures without cryostats and cryogenics, which is a huge advantage in cost and space.

\paragraph{\textbf{Fourier optics}}

A standard micro-photoluminescence setup is constructed by the concept of Fourier optics. It allows us to access exciton-polaritons both in real and reciprocal spaces. In Fourier optics, a lens  transforms an image located at one focal plane to its Fourier transformed image at the opposite focal plane~\cite{Hecht}. Real space in one side of the lens becomes momentum space on the other side of the lens, and vice versa. We implement a compact and powerful setup in real and momentum spaces with different sets of lenses drawn in Fig.~\ref{fig:6}a and b, respectively, utilizing this transformation property. We put a simple charge-coupled detector to record intensities of each pixel at the arrival of photons, a spectrometer to provide energy-resolved spectra and a streak camera for the temporal responses.

\subsubsection{Measurement Setup for Coherence Functions}

R. J. Glauber established a guiding framework in quantum optics, that characterizes statistical properties of the EM fields~\cite{Glauber63}. Normalized coherence functions $g^{(n)} (\vec{r_1}, t_1; \vec{r_2},t_2;...;\vec{r_n},t_n)$ for all orders $n$ are clearly defined. Let us pay a particular attention to the first two coherence functions written by a field operator $\hat{\psi}(\vec{r},t)$ at different positions $(\vec{r_1},\vec{r_2})$  and times $(t_1,t_2)$:

\begin{eqnarray*}
 g^{(1)}(\vec{r_1},t_1; \vec{r_2},t_2) = \frac{\langle \hat{\psi}^{\dagger}(\vec{r_1},t_1)) \hat{\psi}(\vec{r_2},t_2)  \rangle}{\sqrt{ \langle \hat{\psi}^{\dagger}(\vec{r_1},t_1) \hat{\psi}^{\dagger}(\vec{r_1},t_1) \rangle \langle \hat{\psi}^{\dagger}(\vec{r_2},t_2) \hat{\psi}(\vec{r_2},t_2) \rangle}}, \nonumber\\
\end{eqnarray*}
where $\langle \rangle$ indicates the thermal averaging, and

\begin{eqnarray*}
 g^{(2)}(\vec{r_1},t_1; \vec{r_2},t_2) = \frac{\langle \hat{\psi}^{\dagger}(\vec{r_1},t_1)) \hat{\psi}^\dagger(\vec{r_2},t_2) \hat{\psi}^{\dagger}(\vec{r_2},t_2)) \hat{\psi}^\dagger(\vec{r_1},t_1)  \rangle}{ \langle \hat{\psi}^{\dagger}(\vec{r_1},t_1) \hat{\psi}^{\dagger}(\vec{r_1},t_1) \rangle \langle \hat{\psi}^{\dagger}(\vec{r_2},t_2) \hat{\psi}(\vec{r_2},t_2) \rangle}.\nonumber\\
\end{eqnarray*}

It becomes easier to interpret the above formulas when we look at the same position or at the same time.  $g^{(1)}(\vec{r_1}; \vec{r_2})$ tells us the spatial coherence property of the field, a measure of long-range spatial order. $g^{(1)}(t_1; t_2)$ quantifies the temporal coherence property related to the spectral linewidth. When we rewrite the second-order coherence function in terms of the particle density operator $n(\vec{r},t) = \hat{\psi}^{\dagger}(\vec{r},t)\hat{\psi}(\vec{r},t)$ and the variance $\Delta n(\vec{r},t) =  n(\vec{r},t) - \langle n(\vec{r},t) \rangle$, the meaning of $g^{(2)}(\vec{r},t)$ becomes clear,

\begin{eqnarray*}
 g^{(2)}(\vec{r},t; \vec{r},t) = \frac{ \langle n(\vec{r},t)^2 \rangle}{ \langle n(\vec{r},t)\rangle^2} = 1+ \frac{\langle \Delta n(\vec{r},t) ^2 \rangle}{\langle n(\vec{r},t)\rangle^2}. \nonumber\\
\end{eqnarray*}
That is the variance of the number distribution, giving the information of the intensity-intensity correlation.

The optical properties of exciton-polaritons have also been characterized with these first two coherence functions, and we briefly describe standard setups to measure these quantities: an interferometer for the first-order coherence function and a Hanbury Brown-Twiss setup for  the second-order coherence function.

\paragraph{\textbf{Interferometry}}

As mentioned earlier, the condensate order parameter is complex-valued with two real variables, particle density $n(\vec{r})$ and phase $\phi(\vec{r})$. The imaging intensity is proportional to the exciton-polariton density, but the phase information, which is as important as the density, is lost in imaging. For photons,  we construct interferometers as a means to extract this phase information. For exciton-polaritons, Michelson and Mach-Zehnder interferometers (Fig.~\ref{fig:6}c) are used, where the 2D signals interfere with a constant-phase plane wave as reference 2D signal. Analyzing the visibility contrast of this interference image can give a relative phase map of the exciton-polaritons. In combination with the above-mentioned Fourier optics techniques, the interference measurement can be performed both in real and reciprocal spaces. Furthermore, time-resolved and energy-resolved interferograms are also possible together with a spectrometer and a streak camera.

A time-integrated phase map at two different real space coordinates reveals the spatial coherence function $g^{(1)}(\vec{r_1}; \vec{r_2})$, with which we measure the off-diagonal long-range order, a crucial concept of exciton-polariton condensation~\cite{Kasprzak06,Deng07,Roumpos12,Nitsche14}. The temporal coherence function $g^{(1)}(t_1, t_2)$ can be measured using the streak camera, which tells us the dephasing time of the condensates.

\paragraph{\textbf{Hanbury Brown-Twiss Setup}}

A historical Hanbury Brown-Twiss setup~\cite{Brown56} consists of two photon detectors(Fig.~\ref{fig:6}d), which record the number of arrival photons. The data of these detectors are analyzed for the second-order coherence functions. Zero-time delay $g^{2}(0)$ tells us statistical attribute of exciton-polariton condensates. As a well-known fact for a coherent state  $g^{2}(0)$ equals to 1, while it is 2 for a thermal state. Compared to purely coherent or thermal states, $g^{2}(0)$ of the exciton-polariton condensates is found to be between 1 and 2~\cite{Deng02,Horikiri10,Assmann11}; namely exciton-polaritons may not be in a purely coherent nor thermal state.  The deviation from  $g^{2}(0)=1$ in exciton-polaritons would come from polariton-polariton repulsive interactions.

Overall, Table 1 summarizes the exciton-polariton toolbox to obtain quantitative information of both static and dynamic variables, energy and time in real and momentum spaces. 
\begin{table}
\caption{Exciton-Polariton Toolbox.}
\label{tab:1}  
\begin{tabular}{p{4cm}p{4cm}p{3.5cm}}
\hline\noalign{\smallskip}
 & Real Space & Momentum Space  \\
\noalign{\smallskip}\svhline\noalign{\smallskip}
Density & $n(\vec{r}; E; t)$ & $n(\vec{k}; E; t)$ \\
Phase & $\phi(\vec{r}; E; t)$ & $\phi(\vec{k}; E;t)$\\
First order correlation & $g^{(1)}(\vec{r_1},\vec{r_2}; E; t)$ &  $g^{(1)}(\vec{k_1}, \vec{k_2}; E; t)$\\
intensity correlation & $g^{(2)}(\vec{r_1},\vec{r_2}; E; t)$  & $g^{(2)}(\vec{k_1}, \vec{k_2}; E; t)$\\
\noalign{\smallskip}\hline\noalign{\smallskip}
\end{tabular}
\end{table}

\section{Software of Exciton-Polariton Analog Quantum Simulators}
\label{sec:3}

We have discussed that how the exciton-polariton AQS can be physically realized in section 2, reviewing the basics of exciton-polaritons and their condensation properties followed by several ways to engineer exciton-polariton crystals. Then, a collection of measurement setups was explained to assess the optical properties of exciton-polaritons. In this section, we describe applications of the exciton-polariton AQS,  what physical problems can be studied within single-particle and many-particle physics.

\subsection{Single-Particle Physics}
\label{subsec:1}

When we examine a system consisting of many particles in a crystal structure, the first task to know is the band structure of the system, which reflects the topology of the crystal lattices and closely links to the physical properties of the system. It is incredibly difficult to calculate the exact band structures of such a system including all particles as well as their degrees of freedom. Instead of giving up this formidable task, physicists have come up with clever approximations and developed insightful numerical techniques, which illuminate principal nature of the system. A mean-field approximation has been very useful and successful, which reduces the many-body problem into a single-particle physics. It is truly amazing that such a simple single-particle description can faithfully represent the crucial electrical and optical properties of  semiconductors assuming a dielectric constant and an effective mass of constituent particles. Learning from this invaluable lesson, we first evaluate the band structures of exciton-polaritons in artificially patterned lattices with four different 2D geometries.

There are salient phenomena in solid-state materials, which root from the orbital nature of electrons, especially high-orbital electrons with $p$-, $d$- and $f$-wave spatial symmetry~\cite{Tokura00}. Unlike $s$-orbitals, high-orbital wavefunctions share energy degeneracy and exhibit anisotropic distributions in space. Interplaying with spin and charge degrees of freedom, the orbital nature of electrons is responsible for famous phenomena: metal-insulator transition~\cite{Imada98}, colossal magnetoresistance~\cite{Tokura00,Salamon01}, and the newly discovered iron-pnictide superconductors~\cite{Ishida09,Mazin09}. Several schemes were proposed to elucidate orbital physics in bosonic atom-lattice AQS~\cite{Isacsson05,Vincent06}, and orbital states of coherent boson gases were selectively prepared by population transfer from the ground states~\cite{Muller07,Wirth11,Olschlager11,Soltan12}.

We are also captivated by a few fascinating features which fall within single-particle physics. A charged particle in a magnetic field would lead to a quantum Hall effect, resulting from time-reversal symmetry breaking~\cite{Davies}. This physics can be studied in exciton-polariton AQS because the QW exciton is subject to the Zeeman splitting in magnetic fields. Since rotating condensates and the quantum Hall effect  are isomorphic in mathematical models~\cite{Roncaglia11}, exciton-polariton condensates can exhibit quantum Hall physics by rapid rotation. Furthermore, spin-orbital motion of QW excitons can be understood within a single-particle description.

\subsection{Many-Particle Physics}
\label{subsec:2}

Despite the success of single-particle approximations, there are a large number of problems in nature beyond single-particle physics.  Knowing that materials are composed of many atoms, it is impossible to neglect the many-particle physics completely. The simplest example is a two-body interaction, such as a long-ranged Coulomb interaction between two charged particles, which is prevalent in condensed matter systems.  The earlier example, the Hubbard Hamiltonian is widely used to describe properties of transition metal oxides and high-temperature superconductors~\cite{Mahan}.  At the extremes of the ratio of the two energy terms, the system behaves  completely different, either metallic for the stronger kinetic energy case or insulating for the strong interaction case~\cite{Mahan}. 

Motivated by the seminal work in the atom-lattice system~\cite{Greiner02}, Byrnes and colleagues derived the Bose-Hubbard Hamiltonian in exciton-polariton condensates under a 2D periodic potential, identifying a region in phase space to observe superfulid-Mott transitions in exciton-polaritons~\cite{Byrnes10}. Here the on-site interaction term is calculated from exciton-exciton exchange interactions, and the interesting phase transition in 2D is predicted to occur around $U/t \sim 23$, requiring a strong lattice potential of about 6 meV and a small lattice constant on the order of 0.5 $\mu$m. A current challenge to demonstrate this proposed phase transition is to generate a sub-micron trap size to increase the interaction term. Unfortunately, aforementioned methods discussed in section 2.3 are not able to fulfill this stringent requirement. The thin-metal-film technique can lithographically create a trap size as small as 50 nm; however, its potential is too weak, less than 1 meV. If we are able to realize this phase transition, we open a new world such that the Mott-insulating phase would be a basis for generating single-photon arrays on a macroscopic scale~\cite{Na10}, as an initial step towards universal quantum computation.

Ter\c{c}as {\it{et al.}} proposed a scheme of applying a gauge field to a propagating exciton-polariton condensate from the transverse electric and transverse magnetic modes of the cavity and their energy splitting~\cite{Tercas13}. It is a timely research direction to address spin-orbital interactions as a driving force to search for new quantum matter including topological insulators. Recent progress in this direction of research was reported  to observe similar insulating states in photonic systems~\cite{Rechtsman13,Khanikaev13}. The exciton-polariton AQS is  an appropriate platform to investigate these phenomena, and we envision that exciton-exciton interactions would go beyond photonic insulating states, which would be much closer to real material dynamics.  

%
One of the challenging problems is dynamics in open environments or a driven-dissipative situation. The non-equilibrium nature of exciton-polariton systems, unavoidable loss and constant replenishment to compensate loss, may be advantageous to investigate difficult open system dynamics. At present, concrete ideas to address this problem in the exciton-polariton AQS are yet to be developed; however, it is worthwhile defining reachable problems which the exciton-polariton AQS can contribute uniquely well.

\section{Exciton-Polaritons in Two-Dimensional Lattices}
\label{sec:4}

We launched the research project to build an exciton-polariton AQS in 2006, impressed by the atomic AQS in 2002. Section 4 updates the current status of our exciton-polariton AQS based on GaAs semiconductors, and we present experimental results in exciton-polariton-lattice systems. 

\subsection{Experiment}
\label{subsec:1}

\subsubsection{Device}

Our wafer under study was grown by molecular beam epitaxy, which has a superb controllability and fidelity of the layer thickness down to atomic length-scales. An AlAs $\lambda/2$-cavity is capped with DBRs by an alternating layer of Ga$_{0.8}$Al$_{0.2}$As and AlAs. The 16 and 20 pairs of top and bottom DBRs form a planar Fabry-Perot cavity with a quality factor $Q \sim$ 3000, resulting in a cavity photon lifetime of around 2 ps.

At the three antinodes of the EM field in the cavity, a stack of four 7-nm-thick GaAs QWs separated by a 3-nm-thick AlAs barrir are inserted respectively for the purpose of diluting the exciton density per each QW. The whole structure is designed at $\lambda  \sim $ 770 nm, close to the emission wavelength of the 7-nm-thick GaAs QW exciton. The vacuum Rabi splitting of the 12 GaAs QWs is $\sim$15 meV, which was experimentally confirmed from position-dependent reflectance spectra. The wafer also has spatial detunings, which can vary between -15 meV and 15 meV. Most of the experiments were performed with devices at near zero or red detuning areas of the wafer (-3 $\sim$ -5 meV) .

\subsubsection{Lattices}

We prepared exciton-polariton 2D lattice devices with the previously grown GaAs wafer. Three basic (square, triangular and honeycomb) and one complex (kagome) lattices were designed and patterned by electron-beam lithography. A 23/3 nm-Ti/Au think film was deposited as a final step of the fabrication. All of these semiconductor processing steps are considered standard and simple. The center-to-center distance of the nearest neighbor sites are fixed between 2 and 20 $\mu$m. Figure~\ref{fig:7}a presents a photo of three basic lattices in one device. The photon potential depth ranges between 200-400 $\mu$eV, consequently the actual potential strengths for the LPs are around 100-200 $\mu$eV at different detuning positions~\cite{Kim08}. The kinetic energy at the boundaries of the first Brillouin zone (BZ) is around 0.4-1 meV for lattice constants 2-4 $\mu$m. Comparing the potential energy with the kinetic energy, these potentials only weakly perturb the exciton-polaritons.

Owing to weak lattice potentials, justifiably, the band structures of given lattices are calculated using a single-particle plane-wave basis. The complex bands are often displayed along high symmetry points, following the point-group theory. Figure~\ref{fig:7}b and ~\ref{fig:7}c are BZs of square and honeycomb lattices with high symmetry points denoted as $\Gamma, X, M $ and $K(K^\prime)$.  In particular, these points are closely related to the degree of rotational symmetry in the lattice geometries. $\Gamma$ and $M$ in the square lattice satisfy four-fold rotational symmetry, whereas $X(Y)$ has two-fold rotational degeneracies. A weak periodic potential lifts band degeneracies at high symmetry points, and momentum valleys are protected by the gap energy, comparable to the periodic potential strength on the order of 100 $\mu$eV.

\subsubsection{Experimental Setup}

A low-temperature micro-photoluminescence setup was built as a basic characterization tool to study exciton-polaritons in lattices. Since the GaAs QW exciton binding energy is around 10 meV, we kept the device temperatures  below 10 K to avoid thermal dissociation of the excitons. With a replaceable lens, one can select the measurement domain to be either real or reciprocal space. For this sample, we fixed the pump laser energy to 1.61543 eV ( $\sim$767.5 nm) around 6 meV above the LP ground state energy in the linear regime.  3-ps-long laser pulses enter the sample obliquely at $\sim$60 degree ( $\sim$ 7.4 $\times$10$^6$ m$^{-1}$) at a 76 MHz repetition rate.

By changing pump power, one can control the density of injected exciton-polaritons in QWs. Using this basic setup, polariton population distributions were mapped in real and momentum spaces.  A nonlinear density increase in population marks the quantum degeneracy threshold. Lattice potentials modify a single-particle LP dispersion to band structures of LPs. Complex order parameters of condensates in lattices are constructed by interferometry, which can help us to determine relative phases in real space. A homodyne detection scheme is implemented in a Mach-Zehnder configuration. A reference signal at a fixed frequency and a wavenumber is combined against signals of interest. The interference contrast is directly related to the visibility, and the relative phase in real space is extracted.


\subsection{Bosonic Orbital Order}
\label{subsec:2}

\begin{figure}[t]
\sidecaption[t]
\includegraphics[scale=.5]{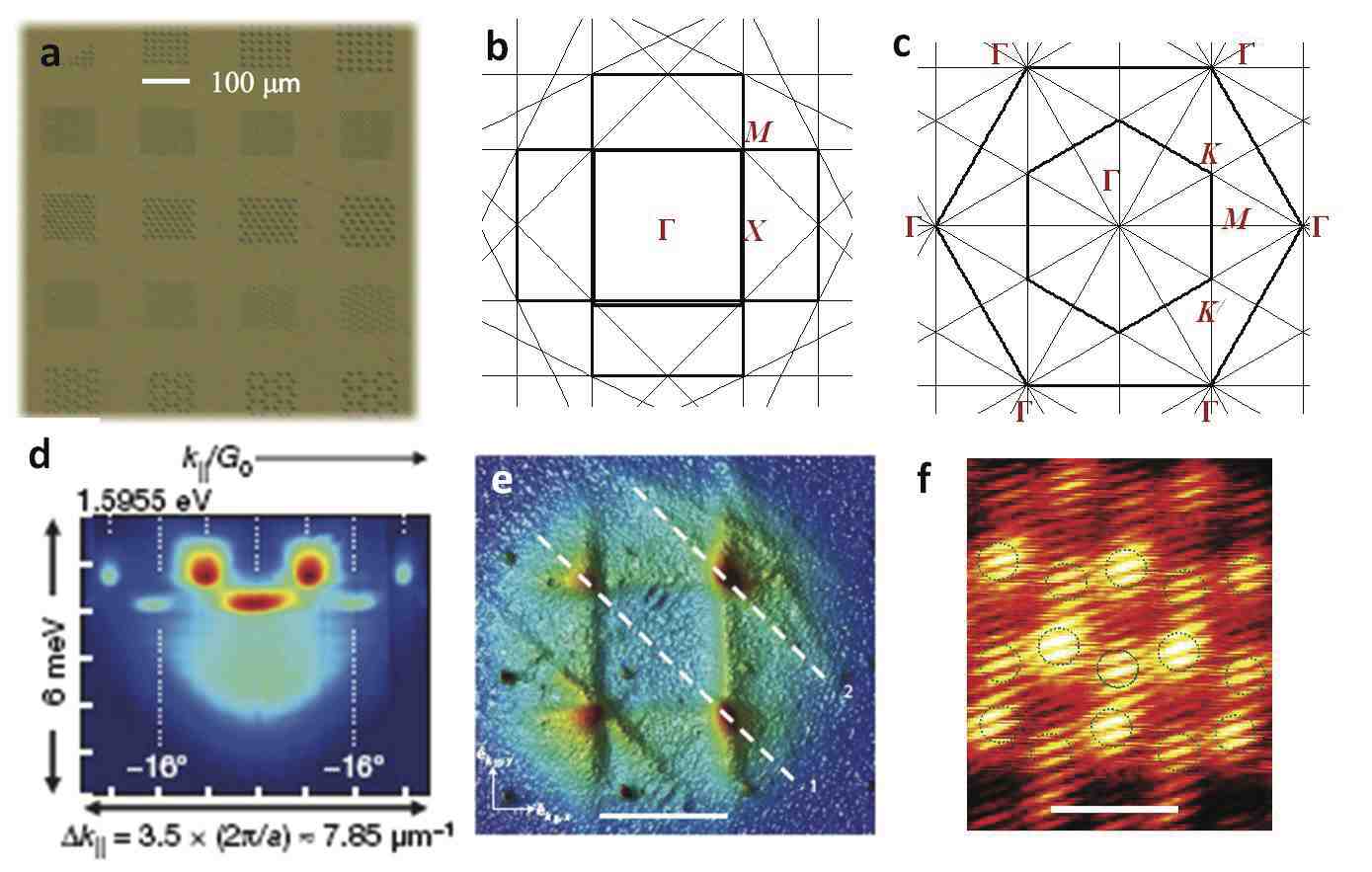}
\caption{\textbf{a}, A photograph of the two-dimensional lattices with square-, triangular- and honeycomb-geometries patterned by photolithography and deposited with a thin Ti/Au metal-film.  Brillouin zones of the square (\textbf{b}) and the honeycomb (\textbf{c}) lattices denoted by high symmetry  $\Gamma$, $M$, $X$, $K$ and $K^\prime$ points according to the rotational symmetry of the geometries.  \textbf{d}, In-phased $s$-orbital (lower energy) and anti-phased coherent $p$-orbital polaritons in a one-dimensional array~\cite{Lai07}. \textbf{e}, A $d_{xy}$-orbital polariton condensation occurring at high-symmetry four $M$ points in the first Brillouin zone (BZ) of the square lattice taken from Ref.~\cite{Kim11}. The white bar is the first BZ size, $2\pi/a$ with a lattice constant $a$ = 4 $\mu$m. \textbf{f}, The real-space interferogram of the $f$-orbital condensation at non-zero $\Gamma$ points in the honeycomb lattice. The white bar is a scale bar of 4 $\mu$m. }
\label{fig:7}       
\end{figure}

The gain-loss dynamics in exciton-polaritons is advantageous to form  condensates beyond zero momentum. Thermal exciton-polaritons are  injected from a high energy and large momentum particle reservoir, and undergo relaxation down through polariton-phonon and polariton-polariton scattering processes.  Since polaritons have a finite lifetime to leak through the cavity, condensation is stabilized at non-zero momentum states in band structures resulting from the balance between two time scales:  the relaxation time to the lower energy states versus the decay time through the cavity. For this very reason, we have observed polariton condensates of high-orbital symmetry: $d_{xy}$-wave condensation at $M$ points in the square lattice, singlet $f$-wave condensations at $\Gamma$ and degenerate $p$-wave condensations at $K$ (or $K^\prime)$ points in both triangle and honeycomb lattices. The degeneracy and anisotropic distribution of high orbital symmetry have been identified with the complex order parameter constructed from interferograms.

The meaning of high-orbital boson order can be found in the following context. We know that  high-orbital electrons play a crucial role in special electrical, chemical and mechanical properties of solid state materials.  $s$-orbital fundamental bosons cannot make any complementary picture to such high-orbital fermionic physics. However, the complex-valued high-orbital bosons can be as important as those of fermions. Along this same spirit, we have observed the anti-phased $p$-orbital condensates in one-dimensional arrays from the diffraction peaks in energy- and angle-resolved spectroscopy shown in Fig.~\ref{fig:7}d~\cite{Lai07}. In 2D lattices, meta-stable momentum valleys are available at high symmetry points. At appropriate pump powers, coherent exciton-polaritons are accumulated at non-zero momentum meta-stable bands.  $d_{xy}$-wave symmetry is favorable at $M$ points of the square lattice, as a signature of narrow coherent diffraction peaks captured in the momentum space image (Fig.~\ref{fig:7}e)~\cite{Kim11}. In hexagonal geometries, 4$f_{y^3-3x^2y}$-like orbital symmetry in space is identified from an interferogram, where alternating phase shifts by $\pi$ at six lobes in real space is unequivocally detected as shown in Fig.~\ref{fig:7}f.

Degenerate $p$-orbital condensates are stabilized at the vertices $K$ and $K^\prime$ points in the first hexagonal BZ of triangular-geometry  lattices. This observation of degenerate condensates raises a fundamental conceptual question of condensation regarding spontaneous symmetry breaking, condensate fragmentation and the statistical mixture of degenerate condensates. In order to address this important question, we have measured $g^{(2)}(t=0)$ in momentum space, which is further described in the following subsection.

\begin{figure}[t]
\sidecaption[t]
\includegraphics[scale=.4]{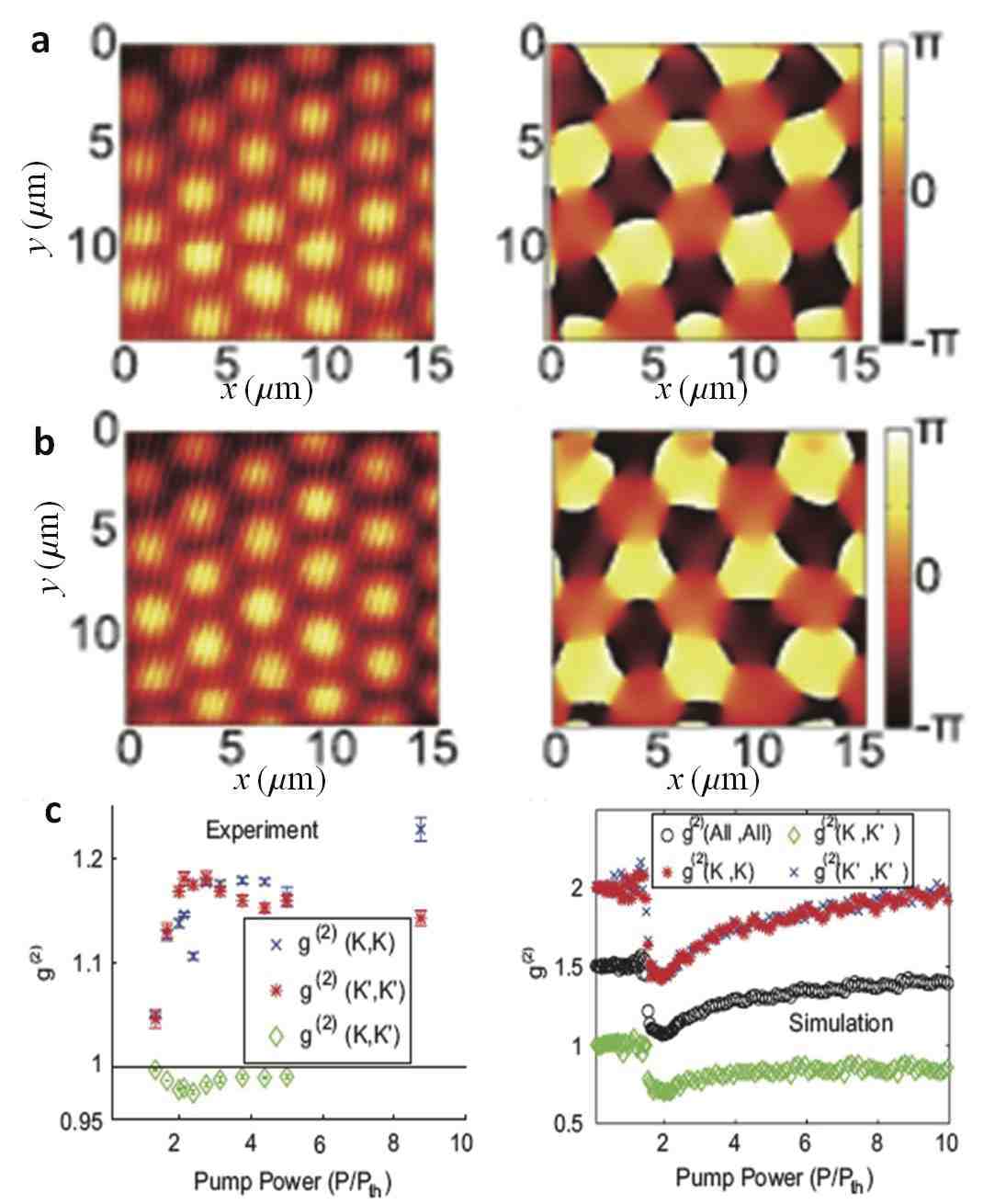}
\caption{$p$-orbital degenerate vortex-antivortex condensates in a honeycomb lattice. Raw interferograms (left) and extracted phase maps (right) to identify a vortex-antivortex single-particle state at $K$ (\textbf{b}) and mirror-symmetric $K^\prime$ (\textbf{c}) points . \textbf{c}, Second-order intensity-intensity auto- and cross-correlation functions of two degenerate $p$-orbital condensates: experimental data (left) and simulation (right).}
\label{fig:8}       
\end{figure}

\subsection{Degenerate High-Orbital Condensates}
\label{subsec:3}

Six vertices of the first BZ in the honeycomb lattice are grouped into inequivalent $K$ and $K^\prime$ points, which are connected with reciprocal lattice vectors. $K$ and $K^\prime$ points hold reflection or inversion symmetry, namely, they are mirror imaged. Three degenerate energy states at $K$ and $K^\prime$ points are split into a high energy singlet and lower energy doublets. The upper singlet state with $p$-orbital symmetry is interesting as a vortex-antivortex state from the linear combination of all three $K$ points, which are rotated by $2\pi/3$~\cite{Kusudo13}. We call vortex to be a topological defect, whose density is zero at the core and whose phase continuously changes by 2$\pi$. An antivortex has the opposite phase-rotation direction. Both vortices and antivortices are located at the trap sites with zero density. The vortex-antivortex order is exactly opposite for $K$ (Fig.~\ref{fig:8}a) and $K^\prime$ (Fig.~\ref{fig:8}a) points revealing the reflection symmetry, confirmed via a modified Mach-Zehnder interferometer.

Next we study the dynamics of the degenerate condensates at $K$ and $K^\prime$ points by measuring intensity correlation functions in momentum space. $g^{(2)}(K,K; \tau=0)$ and $g^{(2)}(K^\prime,K^\prime; \tau=0)$ are the autocorrelation functions at the same $K$ and $K^\prime$ points with zero time delay $\tau=0$. We also detect cross-correlation functions  $g^{(2)}(K,K^\prime; \tau=0)$ between $K$ and $K^\prime$ signals. Experimental results of pump-power dependent correlation functions are presented in Fig.~\ref{fig:8}c (left), and theoretical simulation results based on complex-number Langevin equations are displayed in Fig.~\ref{fig:8}c (right). A value of less than 1 for  $g^{(2)}(K,K^\prime; \tau=0)$ indicates anti-correlation between the two intensities, which we interpret as the mode competition between $K$ and $K^\prime$ condensates stochastically. At first, one state would be selected randomly between two possibilities, then that chosen path would be more favorable due to subsequent stimulated scatterings. At a given pump power, the total particle number is fixed such that a larger intensity in one state links to a smaller intensity in the other, yielding the anticorrelation. At present, we cannot answer interesting physical questions such as spontaneous symmetry breaking or fragmentation of degenerate condensations, primarily because our measurements are limited to be time averaged over many pulses. A single shot measurement will lead us to reach a conclusive answer, whose setup may be available in the near future.

\section{Outlook}
\label{sec:5}

We have given an overview of the exciton-polariton-lattice system for the application of quantum simulators, specially designed functional quantum machines. Physical elements of the exciton-polariton AQS hardware were concretely identified, and the first generation exciton-polariton AQS has already been experimentally prepared and tested. The initial tasks of the exciton-polariton AQS are to directly map the band structures of 2D lattices by means of angle-resolved photoluminescence spectroscopy. We have identified  meta-stable exciton-polariton condensates with high-orbital symmetries at non-zero momentum states. The selectivity of condensate orbital symmetry is controlled by the particle density with pump laser intensity in order to balance lifetime and a relaxation time in the polariton band structures.

The main limitations of the first generation exciton-polariton AQS are a weak potential depth and a micron-sized trap unit, yielding a small gap energy at high symmetry points which is not easily detectable due to a broader spectral linewidth of the generated signals. Hence, how to produce stronger potentials is an imminent goal to be accomplished in order to run interesting software programs. Next, in a micron-sized site, two-particle polariton-polariton interaction is still too weak; estimated value is around tens of $\mu$eV for GaAs QW excitons. It is a primary hindrance to study many-body physics. Unless overcoming this challenge, the initial exciton-polariton AQS can solely investigate problems within single-particle physics, such as quantum Hall effects. Either the direct application of magnetic fields, rapid rotation or generation of a gauge field would be incorporated in this direction of researches.  

As a final remark, we actively seek a class of unique problems the exciton-polariton AQS can simulate much better than other AQS platforms. Appreciating the non-equilibrium nature of the exciton-polariton system, we anticipate that the exciton-polariton AQS may be a promising testbed to examine dynamical problems in open-dissipative environments.  A starting point is to set up a testable toy model through careful and systematic assessment. Witnessing the incredible development of classical computers, we also put our faith in the continuous advancement of quantum simulators in all platforms, which would deepen our knowledge of quantum many-body problems in various areas and would provide crucial insights and novel methods for quantum engineering and technologies. 
\begin{acknowledgement}
We acknowledge Navy/SPAWAR Grant N66001-09-1-2024, the Japan Society for the Promotion of Science (JSPS) through its ``Funding Program for World-Leading Innovative R$\&$D on Science and Technology (FIRST Program)". We deeply thank our collaborators: Dr. K. Kusudo and Dr. N. Masumoto for experimental measurement and device fabrication; Prof. A. Forchel, Dr. S. H\"ofling, Dr. A. L\"offler for providing the wafers;  Prof. T. Fujisawa, Dr. N. Kumada for supporting the device fabrication; Prof. T. Byrnes, Prof. C. Wu, Dr. Z. Cai for theoretical discussions. N.Y.K thank Dr. C. Langrock for critical reading of the manuscript. 
\end{acknowledgement}


\begin{thebibliography}{99.}%

\bibitem{Feynman82} Feynman R.: Simulating physics with computers. Int. J. Theor. Phys. \textbf{21}, 467--488 (1982)
\bibitem{Mahan} Mahan G.D.: Many-particle physics. Kluwer Academic/Plenum Publishers, New York (1981)
\bibitem{Lloyd96} Lloyd S.: Universal Quantum Simulators. Science \textbf{273}, 1073--1078 (1996)
\bibitem{Buluta09} Buluta I., Nori F.: Quantum Simulators. Science \textbf{326}, 108--111 (2009)
\bibitem{Cirac12} Cirac J.I., Zoller P.: Golas and opportunites in quantum simulation. Nat. Phys. \textbf{8}, 264--266 (2012)
\bibitem{Georgescu13} Georgescu I.M., Ashhab S., Nori F.: Quantum Simulation. Rev. Mod. Phys. \textbf{86}, 153--195 (2014)
\bibitem{Ladd10} Ladd T.D., Jelezko F., Laflamme R., Nakamura Y., Monroe C., O'Brien J.L.: Quantum computers. Nature \textbf{464}, 45--53 (2010)
\bibitem{Bloch12} Bloch I., Dalibard J., Nascimb\'ene S.: Quantum simulations with ultracold quantum gases. Nat. Phys. \textbf{8}, 267--276 (2012)
\bibitem{Blatt12} Blatt R., Roos C.F.: Quantum simulations with trapped ions. Nat. Phys. \textbf{8}, 277--284 (2012)
\bibitem{Aspuru12} Aspuru-Guzik A., Walther P.: Photonic quantum simulators. Nat. Phys. \textbf{8}, 285--291 (2012)
\bibitem{Houck12} Houck A.A., T\"ureci H., Koch J.: On-chip quantum simulation with superconducting circuits. Nat. Phys. \textbf{8}, 292--299 (2012)
\bibitem{Lu12} Lu D., Xu B., Xu N., Li Z., Chen H., Peng X., Xu R., Du J.: Quantum chemistry simulation on quantum computers: theories and experiments. Phys. Chem. Chem. Phys. \textbf{14}, 9411--9420 (2012)
\bibitem{Hauke12} Hauke P., Cucchietti F.M., Tagliacozzo L., Deutsch I., Lewenstein M.: Can one trust quantum simulators? Rep. Prog. Phys. \textbf{75}, 082401 (2012)

\bibitem{Greiner02} Greiner M., Mandel O., Esslinger T., H\"ansch T., Bloch I.: Quantum phase transition from a superfulid to a Mott insulator in a gas of ultracold atoms. Nature \textbf{415}, 39--44 (2002) 
\bibitem{Esslinger10} Esslinger T.: Fermi-Hubbard Physics with Atoms in an Optical Lattice. Annu. Rev. Condens. Matter Phys. \textbf{1}, 129--152 (2010) 
\bibitem{Kimk11} Kim K. et al: Quantum simulation of the transverse Ising model. New J. Phys. \textbf{13}, 105003 (2011)
\bibitem{Britton12} Britton J.W., Sawyer B.C., Keith A.C., Joseph Wang C.-C., Freericks J.K., Uys H., Biercuk M.J., Bollinger J.J.: Engineered two-dimensional Ising interactions in a trapped-ion quantum simulator with hundreds of spins. Nature \textbf{484}, 489--492 (2012)
\bibitem{Byrnes06} Byrnes T., Recher P., Kim N.Y., Utsunomiya S., Yamamoto Y.: Quantum simulator for the Hubbard model with long-range Coulomb interactions using surface acoustic waves. Phys. Rev. Lett. \textbf{99}, 016405 (2006)
\bibitem{Byrnes07} Byrnes T., Kim N.Y., Kusudo K., Yamamoto Y.: Quantum simulation of Fermi-Hubbard models in semiconductor quantum dot arrays. Phys. Rev. B \textbf{78}, 075320 (2007)
\bibitem{Simoni10} De Simoni G., Singha A., Gibertini M., Karmakar B., Polini M., Piazza V., Pfeiffer L.N., West K.W., Beltram F., Pellegrini V.: Delocalized-localized transition in a semiconductor two-dimensional honeycomb lattice. Appl. Phys. Lett. \textbf{97}, 132113 (2010)
\bibitem{Singha11} Singha A., Gibertini M., Karmakar B., Yuan S., Polini M., Vignale G., Katsnelson M.I., Pinczuk A., Pfeiffer L.N., West K.W., Pellegrini V.: Two-Dimensional Mott-Hubbard Electrons in an Artificial Honeycomb Lattice. Science \textbf{332}, 1176--1179 (2011)
\bibitem{Koch10} Koch J., Houck A.A., Le Hur K., Girvin S.M.: Time-reversal-symmetry breaking in circuit-QED-based photon lattices. Phys. Rev. A \textbf{82}, 043811 (2010)
\bibitem{Angelakis07} Angelakis D.G., Santos M.F., Bose S.: Photon-blockade-induced Mott transitions and $XY$ spin models in coupled cavity arrays. Phys. Rev. A \textbf{76}, 031805(R) (2007)
\bibitem{Hartmann06} Hartmann M.J., Brand\~{a}o F.G.S.L., Plenio M.B.: Strongly interacting polaritons in coupled arrays of cavities. Nat. Phys. \textbf{2}, 849--855 (2006)
\bibitem{Greentree06} Greentree A.D., Tahan C., Cole J.H., Hollenberg L.C.L: Quantum phase transitions of light. Nat. Phys. \textbf{2}, 856--861 (2006)
\bibitem{Byrnes10} Byrnes T., Recher P., Yamamoto Y.: Mott transitions of excitons polaritons and indirect excitons in a periodic potential. Phys. Rev. B \textbf{81}, 205312 (2010)
\bibitem{Na10} Na N., Yamamoto Y.: Massive parallel generation of indistinguishable single photons iva the polaritonic superfulid to Mott-insulator quantum phase transition. New J.  Phys. \textbf{12}, 123001 (2010)

\bibitem{Hubbard63} Hubbard J.: Electron Correlations in Narrow Energy Bands. Proc. R. Soc. Lond.  A \textbf{276}, 238--257 (1963)
\bibitem{Tokura00} Tokura Y., Nagaosa N.: Orbital Physics in Transition-Metal Oxides. Science \textbf{288}, 462--468 (2000)
\bibitem{Sachdev08} Sachdev S.: Quantum magnetism and criticality. Nat. Phys. \textbf{4}, 173--185 (2008)
\bibitem{Sachdev11} Sachdev S., Keimer B.: Quantum criticality. Phys. Today \textbf{64}(2), 29--35 (2011)

\bibitem{Jaksch03} Jaksch D., Zoller P.: Creation of effective magnetic fields in optical lattices: the Hofstadter butterfly for cold neutral atoms. New J. Phys. \textbf{5}, 56 (2003)
\bibitem{Lin09} Lin Y.-J., Compton R.L., Jim\'enez-Garc\'ia, Porto J.V., Spielman I.B.: Synthetic magnetic fields for ultracold neutral atoms. Nature \textbf{462}, 628--632 (2009)
\bibitem{Galitski13} Glitski V., Spielman I.B.: Spin-orbit coupling in quantum gases. Nature \textbf{494}, 49--54 (2013)

\bibitem{Aspuru05} Aspuru-Guzik A., Dutoi A.D., Love P.J., Head-Gordon M.: Simulated Quantum Computation of Molecular Energies. Science \textbf{309}, 1704--1707 (2005)
\bibitem{Kassal08} Kassal I.S., Jordan S.P., Love P.J., Mohseni M., Aspuru-Guzik A.: Quantum algorithms for the simulation of chemical dynamics. Proc. Nat. Acad. Sci. \textbf{105}, 18681--18686 (2008)
\bibitem{Smirnov07} Smirnov A. Yu., Savel'ev S., Mourokh L.G., Nori F.: Modelling chemical reactions using semiconductor quantum dots. Eur. Phys. Lett. \textbf{80}, 67008 (2007)

\bibitem{Giovanazzi05} Giovanazzi S.: Hawking radiation in sonic black holes. Phys. Rev. Lett. \textbf{94}, 061302 (2005)
\bibitem{Gerace12} Gerace D., Carusotto I.: Analog Hawking radiation from an acoustic black hole in a flowing polariton superfluid. arXiv:1206.4276 (2012)
\bibitem{Gerritsma09} Gerritsma R., Kirchmair G., Z\"ahringer F., Solano E., Blatt R., Roos C.F.: Quantum simulation of the Dirac equation. Nature \textbf{463}, 68--71 (2009)
\bibitem{Tarruell12} Tarruell L., Greif D., Uehlinger T., Jotzu G., Esslinger T.: Creating, moving and merging Dirac points with  Fermi gas in a tunable honeycomb lattice. Nature \textbf{483}, 302--305 (2012)
\bibitem{Gomes11} Gomes K.K., Mar W., Ko W. Guinea F., Manoharan H.C.: Designer Dirac fermions and topological phases in molecular graphene. Nature \textbf{483}, 306--310 (2012)

\bibitem{Weisbuch92} Weisbuch C., Nishioka M., Ishikawa A., Arakawa Y.: Observation of the Coupled exciton-Photon Mode Splitting in a Semiconductor Quantum Microcvity. Phys. Rev. Lett. \textbf{69}, 3314--3317 (1992)
\bibitem{Kavokin} Kavokin, A., Baumberg, J., Malpuech G., Laussy F.P.: Microcavities. Clarendon Press, Oxford (2006)
\bibitem{Snoke10} Snoke D., Littlewood P.: Polariton condensates. Phys. Today  \textbf{63}(8), 42--47 (2010)
\bibitem{Deng10} Deng H., Haug H., Yamamoto Y.: Exciton-polariton Bose-Einstein condensation. Rev. Mod. Phys. \textbf{82}, 1490--1537 (2010)
\bibitem{Yu} Yu P.Y., Cardona M.: Fundamentals of Semciodnuctors. Springer (1996)
\bibitem{Tassone96} Tassone F., Piermarocchi C., Savona V., Quattropani A., Schwendimann P.: Photoluminescence decay times in strong-coupling semiocnductor microcavities. Phys. Rev. B \textbf{53}, R7642--7645 (1996) 
\bibitem{Hanamura77} Hanamura E., Haug H.: Condensation effects of excitons. Phys. Rep. \textbf{33C}, 209--284 (1997)
\bibitem{Griffin} Griffin A., Snoke D.W., Stringari S.: Bose-Einstein Condensation. Cambridge University Press, Cambridge (1995)
\bibitem{Tassone97} Tassone F., Piermarocchi C., Savona V., Quattropani A., Schwendimann P.: Bottleneck effects in the relaxation and photoluminescence of microcavity polaritons. Phys. Rev. B \textbf{56}, 7554--7563 (1997) 
\bibitem{Tassone99} Tassone F., Yamamoto Y. : Exciton-exciton scattering dynamics in a semiconductor microcavity and stimulated scattering into polaritons. Phys. Rev. B \textbf{59}, 10830--10842 (1999) 
\bibitem{Porras02} Porras D., Ciuti C., Baumberg J.J., Tejedor C.: Polariton dynamics and Bose-Einstein condesnation in semiconductor microcavities. Phys. Rev. B \textbf{66}, 085304 (2002) 
\bibitem{Ciuti98} Ciuti C., Savona V., Piermarocchi C., Quattropani A., Schwendimann P.: Role of the exchange of carriers in elastic exciton-exciton scattering in quantum wells. Phys. Rev. B \textbf{58}, 7926--7933 (1998) 
\bibitem{Imamoglu96} Imamoglu A., Ram R.J., Pau S., Yamamoto Y.: Nonequilibrium condensates and lasers without inversion: Exciton-polariton lasers. Phys. Rev. A \textbf{53}, 4250--4253 (1996)
\bibitem{Deng02} Deng H., Weihs G., Santori C., Bloch J., Yamamoto Y.: Condensation of Semiconductor Microcavity Exciton Polaritons. Science \textbf{298}, 199--202 (2002)
\bibitem{Kasprzak06} Kapsrzak J. et al: Bose-Einstein condensation of exciton polaritons. Nature \textbf{443}, 409--414 (2006)
\bibitem{Deng06} Deng H., Press D., G\"otzinger S., Solomon G.S., Hey R., Ploog K.H., Yamamoto Y.: Quantum Degenerate Exciton-Polaritons in Thermal Equilibrium. Phys. Rev. Lett. \textbf{97}, 146402 (2006)
\bibitem{Balili07} Balili R.B., Hartwell V.,  Snoke D., Pfeiffer L., West K.: Bose-Einstein Condensation of Microcavity Polaritons in a Trap. Science \textbf{316}, 1007--1010 (2007)
\bibitem{Deng07} Deng H., Solomon G.S., Hey R., Ploog, K. H., Yamamoto Y.: Spatial Coherence of a Polariton Condensate. Phys. Rev. Lett. \textbf{99}, 126403 (2007)
\bibitem{Chrisopoulos07} Christopoulos S. et al.: Room-Temperature Polariton Lasing in Semiconductor Microcavities. Phys. Rev. Lett. \textbf{98}, 126405 (2007)
\bibitem{Cohen10} K\'ena-Cohen S., Forrest S.R.: Room-temperature polariton lasing in an organic single-crystal microcavity. Nat. Photon.\textbf{4}, 371--375 (2010)
\bibitem{Plumhof14} Plumhof J.D., St\"oferle T., Mai L., Scherf U., Mahrt R.: Room-temperature Bose-Eistein condensation of cavity exciton-polaritons in a polymer. Nat. Mat. \textbf{13}, 247--252 (2014)
\bibitem{Lai07} Lai C. W. et al: Coherent zero-state and $\pi$-state in an exciton-poalriton  condensate array. Nature \textbf{450}, 529--533 (2007)

\bibitem{Balili06} Balili R.B., Snoke D.W., Pfeiffer L., West K.: Actively tuned and spatially trapped polaritons. Appl. Phys. Lett. \textbf{88}, 031110 (2006) 
\bibitem{Miller84} Miller D.A.B., Chemla D.S., Damen T.C., Gossard A.C., Wiegmann W., Wood T.H., Burrus C.A.: Band-Edge Electroabsorption in Quantum Well Structures: The Quantum-Confined Stark Effect. Phys. Rev. Lett. \textbf{53}, 2173--2176 (1984)
\bibitem{deLima06} De Lima M.M.Jr., van der Poel M., Santos P.V., Hvam J.M.: Phonon-Induced Polariton Superlattices. Phys. Rev. Lett. \textbf{97}, 045501 (2006) 
\bibitem{Cerda10}  Cerda-M\'{e}ndez E. A. et al: Polariton Condensation in Dynamic Acoustic Lattices. Phys. Rev. Lett. \textbf{105}, 116402 (2010) 
\bibitem{Cerda13}  Cerda-M\'{e}ndez E. A., Sarkar D., Krizhanovskii D.N., Gavrilov S.S., Biermann K., Skolnick M.S., Santos P.V.: Exciton-Polariton Gap Solitons in Two-Dimensional Lattices. Phys. Rev. Lett. \textbf{111}, 146401 (2013) 
\bibitem{ElDaif06} El Da\"if O. et al: Polariton quantum boxes in semiconductor microcvities. Appl. Phys. Lett. \textbf{88}, 061105 (2006) %
\bibitem{Nardin10} Nardin G., L\'{e}ger Y., Pietka B. Morier-Genoud F., Deveaud-Pl\'{e}dran B. : Phase-resolved imaging of confined exciton-poalriton wave functions in elliptical traps. Phys. Rev. B \textbf{82}, 045304 (2010) 
\bibitem{Bloch98} Bloch J., Boeuf F., G\'{e}rard J.M, Legrand B., Marzin J.Y., Planel R., Thierry-Mieg V., Costard. E.: Strong and weak coupling regime in pillar semiconductor microcavities. Physica E \textbf{2}, 915 (1998) 
\bibitem{Galbiati12} Galbiati M., Ferrier L., Solynshkov D.D., Tanese D., Wertz E., Senellart P., Sagnes I., Lema\^{i}tre, Galopin El, Malpuech G., Bloch J.: Polariton Condensation in Photonic Molecules. Phys. Rev. Lett. \textbf{108}, 126403 (2012) 
\bibitem{Wertz10} Wertz E. et al: Spontaneous formation and optical manipulation of extended polariton condensates. Nat. Phys. \textbf{6}, 860--864 (2010)
\bibitem{Jacqmin14} Jacqmin T. et al: Direct observation of Dirac cones and a flatband in a honeycomb lattice for polartions. Phys. Rev. Lett. \textbf{112}, 116402 (2014) 
\bibitem{Roumpos10} Roumpos G., Nitsche W.H., H\"ofling S., Forchel A., Yamamoto Y.: Gain-Induced Trapping of Microcavity Exciton Polariton Condensates. Phys. Rev. Lett. \textbf{104}, 126403 (2010) 
\bibitem{Tosi12} Tosi G., Christmann G., Berloff N. G., Tsotsis P., Gao T., Hatzopoulos Z., Lagoudakis P.G., Baumberg J. J.: Geometrically locked vortex lattices in semiconductor quantum fluids. Nat. Comms. \textbf{3}, 1243 (2012)
\bibitem{Kim08} Kim N.Y. et al: GaAs microcavity exciton-polaritons in a trap. Phys. Rev. Lett. \textbf{105}, 116402 (2010) 
\bibitem{Kim11} Kim N.Y. et al: Dynamical $d$-wave condensation of exciton-polaritons in a two-dimensional square-lattice potential. Nat. Phys. \textbf{7}, 681--686 (2011)
\bibitem{Masumoto12} Masumoto N., Kim N.Y., Byrnes T., Kenichiro K., L\"offler A., H\"ofling S., Forchel A., Yamamoto Y.: Exciton-poalriton condensates with flat banda in a two-dimensional kagome lattice. New J. Phys. \textbf{14}, 065002 (2012)
\bibitem{Kim13} Kim N.Y., Kenichiro K., L\"offler A., H\"ofling S., Forchel A., Yamamoto Y.: Exciton-poalriton condensates near the Dirac point in a triangular lattice. New J. Phys. \textbf{15}, 035032 (2013)
\bibitem{Kusudo13} Kusudo K., Kim N.Y., L\"offler A., H\"ofling S., Forchel A., Yamamoto Y.: Stochastic formation of polariton condensates in two degenerate orbital states. Phys. Rev. B \textbf{87}, 214503 (2013)
\bibitem{Hecht} Hecht, E.: Optics. Addison-Wesley (2001)
\bibitem{Glauber63} Glauber, R. J.: The Quantum Theory of Optical Coherence. Phys. Rev. \textbf{130}, 2529--2539 (1963)
\bibitem{Roumpos12} Roumpos G., Lohse M., Nitsche W.H., Keeling J., Szymanska M.H., Littlewood P.B., L\"offler A., H\"ofling S., Worschech L., Forchel A., Yamamoto Y.: Power-law decay of the spatial correlation function in exciton-polariton condensates. Proc. Nat. Acad. Sci. \textbf{109}, 6467-6472 (2012)
\bibitem{Nitsche14} Nitsche W.H., Kim N.Y., Roumpos G., Schneider C., Kamp M., H\"ofling S., Forchel A., Yamamoto Y.: Algebraic order and the Berezinskii-Kosterlitz-Thouless transition in an exciton-polariton gas. arXiv:1401.0756 (2014)

\bibitem{Brown56} Hanburry Brown R., Twiss R. Q.: The Test of a New Type of Stella Interferometer on Sirrus. Nature \textbf{177}, 27--29 (1956)
\bibitem{Horikiri10} Horikiri T., Schwendimann P., Quattropani A., H\"ofling S., Forchel A., Yamamoto Y.: Higher order coherence of exciton-polariton condensates. Phys. Rev. B \textbf{81}, 033307 (2010)
\bibitem{Assmann11} A$\ss$mann M. et al: From polariton condensates to highly photonic quantum degenerate states of bosonic matter. Proc. Nat. Acad. Sci. \textbf{108},1804--1809 (2011)

\bibitem{Imada98} Imada M., Fujimori A., Tokura Y.: Metal-insulator transition. Rev. Mod. Phys. \textbf{70}, 1039--1263 (1998)
\bibitem{Salamon01} Salamon M.B., Jaime M.: The physics of manganites: Structure and transport. Rev. Mod. Phys. \textbf{73},  583--628 (2001)
\bibitem{Ishida09} Ishida K., Nakai Y., Hosono  H.: To what extent iron-pnictide new superconductors have been clarifies: A progress report. J. Phys. Soc. Jpn. \textbf{78},  062001 (2009)
\bibitem{Mazin09} Mazin I. I., Schmalian J.: Pairing symmetry and pairing state in ferropnictides:Theoretical overview. Physica C \textbf{469},  614--627 (2009)

\bibitem{Isacsson05} Isacsson A., Girvin S.M.: Multiflavor bosonic Hubbard models in the first excite Bloch band of an optical lattice. Phys. Rev. A \textbf{72}, 053604 (2005)
\bibitem{Vincent06} Vincent Liu W., Wu C.: Atomic matter of nonzero-momentum Bose-Einstin condensation and orbital current order. Phys. Rev. A \textbf{74}, 013607 (2006)
\bibitem{Muller07} M\"uller T., F\"olling S., Widera A., Bloch I: State separation and dynmics of ultracold atoms in higher lattice orbitals. Phys. Rev. Lett. \textbf{99}, 200405 (2007)
\bibitem{Wirth11} Wirth G., \"Olschl\"ager, Hemmercih A.: Evidence for orbital superfluidity in the $P$-band of a bipartite optical square lattice. Nat. Phys. \textbf{7}, 147--153 (2011)
\bibitem{Olschlager11} \"Olschl\"ager M., Wirth G., Hemmercih A.: Unconventional Superfluid Order in the $F$ Band of a Bipartite Optical Square Lattice. Phys. Rev. Lett. \textbf{106}, 015302 (2011)
\bibitem{Soltan12} Soltan-Panahi P., L\"uhmann D., Struck J., Windpassinger P., Sengstock K.: Quantum phase transition to unconventional multi-orbital superfluidity in optical lattices. Nat. Phys. \textbf{8}, 71--75 (2012)
\bibitem{Davies} Davies J.H.: The Physics of Low-dimensioanl Semiconductors: An Introduction. Cambridge University Press, Cambridge (1997)
\bibitem{Roncaglia11} Roncaglia M., Rizzi M., Dalibard J.: From rotating atomic rings to quantum Hall states. Sci. Rep. \textbf{1}, 43 (2011)

\bibitem{Tercas13} Ter\c{c}as H., Flayac H., Solnyshkov D.D., Malpuech G.: Non-Abelian Gauge Fields in Photonic Cavities. Phys. Rev. Lett. \textbf{112}, 066402 (2014)
\bibitem{Rechtsman13} Rechtsman M.C., Zeuner J.M., Plotnik Y., Lumer Y., Podolsky D., Dreisow F., Nolte S., Segev M., Szameit A.: Photonic Floquet topological insulators. Nature \textbf{496}, 196--200 (2013)
\bibitem{Khanikaev13} Khanikaev A.B., Hossein Mousavi S., Tse W.-K., Kargarian M., MacDonald A.H., Shvets G.: Photonic topological insulators. Nat. Mat. \textbf{12}, 233--239 (2013)
    

















%
%
%
%
%
%
%

%
%
%
%
%
%

\end{thebibliography}
\end{document}